\documentclass[twocolumn,english,showpacs,twocolumn,superscriptaddress,nofootinbib,prc]{revtex4-1}
\usepackage[T1]{fontenc}
\usepackage[latin9]{inputenc}
\usepackage{color}
\usepackage{ulem}
\usepackage{babel}
\usepackage{booktabs}
\usepackage{bm}
\usepackage{amsmath}
\usepackage{amssymb}
\usepackage{graphicx}
\usepackage[unicode=true,pdfusetitle,
 bookmarks=true,bookmarksnumbered=false,bookmarksopen=false,
 breaklinks=false,pdfborder={0 0 1},backref=false,colorlinks=true]
 {hyperref}
\hypersetup{ citecolor=blue,linkcolor=blue}

\makeatletter

\providecommand{\tabularnewline}{\\}

\makeatother

\begin{document}
\title{Strange hadron production in a quark combination model in Au+Au collisions
at energies available at the BNL Relativistic Heavy Ion Collider}
\author{Jun Song }
\affiliation{Department of Physics, Jining University, Shandong 273155, China}
\author{Xiao-feng Wang}
\affiliation{School of Physics and Physical Engineering, Qufu Normal University, Shandong
273165, China}
\author{Hai-hong Li}
\affiliation{Department of Physics, Jining University, Shandong 273155, China}
\author{Rui-qin Wang}
\affiliation{School of Physics and Physical Engineering, Qufu Normal University,
Shandong 273165, China}
\author{Feng-lan Shao}
\email{shaofl@mail.sdu.edu.cn}

\affiliation{School of Physics and Physical Engineering, Qufu Normal University,
Shandong 273165, China}
\begin{abstract}
We apply a quark combination model to study yield densities and transverse
momentum ($p_{T}$) spectra of strange (anti-)hadrons at mid-rapidity
in central Au+Au collisions at $\sqrt{s_{NN}}=$ 7.7, 11.5, 19.6,
27, 39 and 200 GeV. We show that the experimental data for $p_{T}$ spectra
of (anti-)hadrons in these collisions can be systematically described
by the equal velocity combination of constituent quarks and antiquarks
at hadronization. We obtain the $p_{T}$ spectra of quarks and antiquarks
at hadronization and study their collision energy dependence. We also
reproduce the yield densities of hadrons and anti-hadrons. In particular,
we demonstrate that the yield ratios of anti-hadrons to hadrons $K^{-}/K^{+}$,
$\bar{p}/p$, $\bar{\Lambda}/\Lambda$, $\bar{\Xi}^{+}/\Xi^{-}$ and
$\bar{\Omega}^{+}/\Omega^{-}$ simply correlate with each other and
their experimental data except $\bar{\Omega}^{+}/\Omega^{-}$ at $\sqrt{s_{NN}}=$
7.7 GeV are systematically described by the model. 
These results suggest that the equal velocity combination mechanism for quarks and antiquarks at hadronization plays an important role for the production of these long-lived hadrons in Au+Au collisions at low RHIC energies ($\sqrt{s_{NN}}\geq$ 11.5 GeV). 
\end{abstract}
\maketitle

\section{Introduction\label{sec:Intro}}

Hadron production from final state partons in high energy collisions
is a complex Quantum Chromodynamics (QCD) process. Due to the difficulty
of non-perturbative QCD, phenomenological mechanisms and models have
to be applied to describe the production of hadrons at hadronization
\cite{Andersson:1983ia,Webber:1983if,Das:1977cp,Anisovich:1972pq,Bjorken:1973mh,Xie:1988wi,Becattini:1995if,BraunMunzinger:1995bp}.
In relativistic heavy-ion collisions at high RHIC and LHC energies,
quark-gluon plasma (QGP) is created in early stage of collisions and
the hadronization of QGP can be microscopically described by the quark
(re-)combination/coalescence mechanism \cite{Greco:2003xt,Fries:2003vb,Molnar:2003ff,Lin:2003jy,Shao:2004cn,Chen:2006vc}.
The enhanced ratio of baryon to meson and number of constituent quark
scaling for elliptic flows of hadrons at the intermediate transverse
momentum ($p_{T}$) are typical experimental signals for quark combination
mechanism at hadronization and have been widely observed in relativistic
heavy-ion collisions \cite{Adare:2006ti,Adamczyk:2015ukd,Abelev:2014pua,Acharya:2018zuq,Adamczyk:2013gw,Adcox:2001mf,Abelev:2006jr,Abelev:2013xaa}. 

In our recent studies in high-multiplicity events in $pp$ and $p$-Pb
collisions at LHC energies where the mini-QGP is possibly created
and re-scattering of hadrons is weak, we found an interesting quark
number scaling property for $p_{T}$ spectra of hadrons at mid-rapidity
\cite{Song:2017gcz,Zhang:2018vyr}. This scaling property is a direct
consequence of equal velocity combination (EVC) of constituent quarks
and antiquarks at hadronization. Our studies showed that the EVC of
up/down, strange and charm quarks can provide a good and systematic
description on $p_{T}$ spectra of light, strange and charm hadrons
in ground-state in $pp$ and $p$-Pb collisions at LHC energies \cite{Song:2017gcz,Gou:2017foe,Zhang:2018vyr,Song:2018tpv,Li:2017zuj}.
In latest work \cite{Song:2019sez}, we further found that the experimental
data for $p_{T}$ spectra of $\Omega$ and $\phi$ at mid-rapidity
in heavy-ion collisions in a broad collision energy range ($\sqrt{s_{NN}}=11.5-2760$
GeV) also satisfy the quark number scaling property. This is a clear
signal of EVC at hadronization even in heavy-ion collisions. Therefore,
it is interesting to systematically test this mechanism by production
of more hadron species in relativistic heavy-ion collisions. 

Recently, STAR collaboration reported their precise experimental data
for the production of strange hadrons in Au+Au collisions at $\sqrt{s_{NN}}=$7.7-39
GeV \cite{Adam:2019koz}. This provides us a good opportunity to
systematically study the EVC mechanism of hadron production in these
collisions. In this paper, we apply a quark combination model with
EVC to carry out a systematic study on yield densities and $p_{T}$
spectra of strange hadrons in Au+Au collisions at $\sqrt{s_{NN}}=$
7.7, 11.5, 19.6, 27, 39 and 200 GeV. We put particular emphasis on
the self-consistency of the model in explaining the experimental data
for different kinds of hadrons and on the regularity in multi-hadron
production correlations which is sensitive to hadronization mechanism.
Taking advantage of precise data for strange hadrons, we also study
the strangeness neutralization in the midrapidity region in these
collisions. We extract quark $p_{T}$ distributions at hadronization
from data of hadrons and study properties of the relative abundance
for strange quarks as the function of collision energy. Furthermore,
we discuss the key physics in current quark combination model which
are responsible for explaining successfully experimental data of hadronic
$p_{T}$ spectra and yields, and discuss the creation of QGP or the
deconfinement at low RHIC energies.

The paper is organized as follows. In Sec. \ref{sec:sdqcm}, we introduce
a quark combination model with equal velocity combination approximation.
In Sec. \ref{sec:Sconservation}, we study the strangeness neutralization
in mid-rapidity region. In Sec. \ref{sec:hpt}, we show results of
$p_{T}$ spectra for hadrons and compare them with experimental data.
In Sec. \ref{sec:Yields}, we study the split in yield between hadrons
and their antiparticles. In Sec. \ref{sec:quark_prop}, we study properties
for the obtained numbers and $p_{T}$ spectra of quarks at hadronization
at different collision energies. The summary and discussion are given
at last in Sec. \ref{sec:Summary}. 

\section{A quark combination model with EVC \label{sec:sdqcm}}

The quark combination is one of phenomenological mechanisms for hadron
production at hadronization. The basic idea of quark combination was
firstly proposed in 1970s \cite{Anisovich:1972pq,Bjorken:1973mh,Das:1977cp}
and has many successful applications in high energy reactions \cite{Xie:1988wi,Buschbeck:1980ug,Liang:1991ya,Braaten:2002yt}.
In relativistic heavy-ion collisions, quark combination mechanism
is often used to describe the hadron production at QGP hadronization
\cite{Zimanyi:1999py,Hwa:2002tu,Greco:2003xt,Fries:2003vb,Molnar:2003ff,Shao:2004cn,Chen:2006vc}.
Recently, we found that quark combination can also well explain experimental
data of hadron production in high-multiplicity $pp$ and $p$-Pb collisions
at LHC energies \cite{Song:2017gcz,Gou:2017foe,Shao:2017eok,Zhang:2018vyr}.

In this paper, inspired by the quark number scaling property of hadronic
$p_{T}$ spectra \cite{Song:2017gcz,Zhang:2018vyr,Song:2019sez},
we adopt a specific version of quark combination model \cite{Song:2017gcz,Gou:2017foe}.
This model assumes the combination of constituent quarks and antiquarks
with equal velocity to form baryons and mesons at hadronization. The
unknown non-perturbative dynamics at hadronization are parameterized
and their values are assumed to be stable in high energy reactions
and are fixed by experimental data. This model is essentially a statistical
model based on the constituent quark degrees of freedom at hadronization
and the constituent quark structure of hadrons. It is different from
the popular re-combination/coalescence models which adopt the Wigner
wave function method under instantaneous hadronization approximation
\cite{Fries:2003vb,Greco:2003xt}. The brief description of the model
is as follows.

We start from the general formula for the production of the baryon
$B_{j}$ composed of $q_{1}q_{2}q_{3}$ and the production of the
meson $M_{j}$ composed of $q_{1}\bar{q}_{2}$ in quark combination
mechanism 

\begin{align}
f_{B_{j}}(p_{B}) & =\int dp_{1}dp_{2}dp_{3}{\cal R}_{B_{j}}(p_{1},p_{2},p_{3};p_{B})\label{eq:fb}\\
 & \,\,\,\,\,\,\,\,\times f_{q_{1}q_{2}q_{3}}(p_{1},p_{2},p_{3}),\nonumber \\
f_{M_{j}}(p_{M}) & =\int dp_{1}dp_{2}{\cal R}_{M_{j}}(p_{1},p_{2};p_{M})f_{q_{1}\bar{q}_{2}}(p_{1},p_{2}).\label{eq:fm}
\end{align}
Here $f_{q_{1}q_{2}q_{3}}(p_{1},p_{2},p_{3})$ is the joint momentum
distribution for $q_{1}$, $q_{2}$ and $q_{3}$. The combination
kernel function ${\cal R}_{B_{j}}(p_{1},p_{2},p_{3};p_{B})$ denotes
the probability density for a given $q_{1}q_{2}q_{3}$ with momenta $p_{1},p_{2}$
and $p_{3}$ to combine into a baryon $B_{j}$ with momentum $p_{B}$.
It is similar for mesons. We emphasize that Eqs. (\ref{eq:fb}) and
(\ref{eq:fm}) are generally suitable for hadron production in momentum
space of any dimension. In this paper, we study the one-dimensional
transverse momentum ($p_{T}$) distribution of hadrons at midrapidity
$y=0$. In this case, $p_{i}$ simply denotes $p_{T,i}$ and the distribution
function $f_{h}$$\left(p\right)$ denotes the $dN_{h}/dp_{T}$ at
midrapidity. 

The combination functions ${\cal R}_{B_{j}}(p_{1},p_{2},p_{3};p_{B})$
and ${\cal R}_{M_{j}}(p_{1},p_{2};p_{M})$ contain the key information
of combination dynamics which is not clear at present due to the non-perturbative
difficulty of hadronization. In our recent works \cite{Song:2017gcz,Zhang:2018vyr,Song:2019sez},
we observed an interesting quark number scaling property for the $p_{T}$
spectra of hadrons in high-multiplicity $pp$ and $p$-Pb collisions
as well as in heavy-ion collisions. This scaling property supports
the combination of constituent quarks and antiquarks with equal velocity.
This suggests an effective form for the combination kernel functions,
i.e., 
\begin{align}
{\cal R}_{B_{j}}(p_{1},p_{2},p_{3};p_{B}) & =\kappa_{B_{j}}\prod_{i=1}^{3}\delta(p_{i}-x_{i}p_{B}),\label{eq:RB}\\
{\cal R}_{M_{j}}(p_{1},p_{2};p_{M}) & =\kappa_{M_{j}}\prod_{i=1}^{2}\delta(p_{i}-x_{i}p_{M}).\label{eq:RM}
\end{align}
Here, $\kappa_{B_{j}}$ and $\kappa_{M_{j}}$ are coefficients independent
of momentum. Momentum fraction $x_{i}$ is determined by the masses of constituent
quarks because $p_{i}=m_{i}\gamma\beta\propto m_{i}$. Specifically,
we have $x_{i}=m_{i}/(m_{1}+m_{2}+m_{3})$ for baryon $B_{j}$ with
$x_{1}+x_{2}+x_{3}=1$ and $x_{i}=m_{i}/(m_{1}+m_{2})$ for meson
$M_{j}$ with $x_{1}+x_{2}=1$. $m_{i}$ is the constituent mass for
quark of flavor $i$. Because the mass of the formed hadron under Eqs.~(\ref{eq:RB}) and (\ref{eq:RM}) is the sum of these of constituent quarks, we take contituent masses $m_{u}=m_{d}=0.3$ GeV and $m_{s}=0.5$ GeV in order to properly describe the production of baryons and vector mesons studied in this paper.  

The joint momentum distributions $f_{q_1q_2q_3}(p_1,p_2,p_3)$ and $f_{q_1 \bar{q}_2}(p_1,p_2)$ generally contain the correlation term caused by, for example, the collective flow formed in system evolution before hadronization in heavy-ion collisions. In order to obtain analytical and simple expressions for $f_{B_j}(p_B)$ and $f_{M_j}(p_M)$,
we take the independent distribution approximation
\begin{align}
f_{q_{1}q_{2}q_{3}}(p_{1},p_{2},p_{3}) & =f_{q_{1}}(p_{1})f_{q_{2}}(p_{2})f_{q_{3}}(p_{3}),\label{eq:fqqq}\\
f_{q_{1}\bar{q}_{2}}(p_{1},p_{2}) & =f_{q_{1}}(p_{1})f_{\bar{q}_{2}}(p_{2}).\label{eq:fqqbar}
\end{align}
Substituting Eqs. (\ref{eq:RB})-(\ref{eq:fqqbar}) into Eqs. (\ref{eq:fb})
and (\ref{eq:fm}), we obtain 
\begin{align}
f_{B_{j}}(p_{B}) & =\kappa_{B_{j}}f_{q_{1}}(x_{1}p_{B})f_{q_{2}}(x_{2}p_{B})f_{q_{3}}(x_{3}p_{B}),\label{eq:fbfinal}\\
f_{M_{j}}(p_{M}) & =\kappa_{M_{j}}f_{q_{1}}(x_{1}p_{M})f_{\bar{q}_{2}}(x_{2}p_{M}).\label{eq:fmfinal}
\end{align}

$\kappa_{B_{j}}$ and $\kappa_{M_{j}}$ carry the information of $p_{T}$-independent
combination dynamics. In order to determine their forms, we express
momentum distributions of hadrons in another form 

\begin{align}
f_{B_{j}}\left(p_{B}\right) & =N_{B_{j}}\,f_{B_{j}}^{\left(n\right)}\left(p_{B}\right),\label{eq:fbi2}\\
f_{M_{j}}\left(p_{M}\right) & =N_{M_{j}}\,f_{M_{j}}^{\left(n\right)}\left(p_{M}\right).\label{eq:fmi2}
\end{align}
 $N_{B_{j}}$ and $N_{M_{j}}$ are numbers of $B_{j}$ and $M_{j}$,
respectively. $f_{B_{j}}^{\left(n\right)}\left(p_{B}\right)$ and
$f_{M_{j}}^{\left(n\right)}\left(p_{M}\right)$ are distribution functions
normalized to one when integrating over momentum, which can be obtained
by those of quarks and antiquarks, 

\begin{align}
f_{B_{j}}^{\left(n\right)}\left(p_{B}\right) & =A_{B_{j}}\,f_{q_{1}}^{\left(n\right)}\left(x_{1}p_{B}\right)f_{q_{2}}^{\left(n\right)}\left(x_{2}p_{B}\right)f_{q_{3}}^{\left(n\right)}\left(x_{3}p_{B}\right),\label{eq:fnbi}\\
f_{M_{j}}^{\left(n\right)}\left(p_{M}\right) & =A_{M_{j}}f_{q_{1}}^{\left(n\right)}\left(x_{1}p_{M}\right)f_{\bar{q}_{2}}^{\left(n\right)}\left(x_{2}p_{M}\right).\label{eq:fnmi}
\end{align}
Here, $f_{q_{i}}^{\left(n\right)}\left(p\right)=f_{q_{i}}\left(p\right)/N_{q_{i}}$
is the normalized distribution function of quark $q_{i}$. Normalization
coefficients $A_{B_{j}}$ and $A_{M_{j}}$ are defined as
\begin{align}
A_{B_{j}}^{-1} & =\int f_{q_{1}}^{\left(n\right)}\left(x_{1}p_{B}\right)f_{q_{2}}^{\left(n\right)}\left(x_{2}p_{B}\right)f_{q_{3}}^{\left(n\right)}\left(x_{3}p_{B}\right)dp_{B},\label{eq:ABj}\\
A_{M_{j}}^{-1} & =\int f_{q_{1}}^{\left(n\right)}\left(x_{1}p_{M}\right)f_{\bar{q}_{2}}^{\left(n\right)}\left(x_{2}p_{M}\right)dp_{M}.\label{eq:AMj}
\end{align}

Substituting Eqs. (\ref{eq:fnbi}) and (\ref{eq:fnmi}) into Eqs.
(\ref{eq:fbi2}) and (\ref{eq:fmi2}) and then comparing them with
Eqs. (\ref{eq:fbfinal}) and (\ref{eq:fmfinal}), we obtain 
\begin{align}
N_{B_{j}} & =N_{q_{1}}N_{q_{2}}N_{q_{3}}\frac{\kappa_{B_{j}}}{A_{B_{j}}},\\
N_{M_{j}} & =N_{q_{1}}N_{\bar{q}_{2}}\frac{\kappa_{M_{j}}}{A_{M_{j}}}.
\end{align}
 From above two equations, we can read out the physical meaning of
$\kappa_{B_{j}}$ and $\kappa_{M_{j}}$. It is obvious that $\kappa_{B_{j}}/A_{B_{j}}$
denotes the momentum-integrated probability of $q_{1}q_{2}q_{3}$
forming a baryon $B_{j}$ and $\kappa_{M_{j}}/A_{M_{j}}$ denotes
that of $q_{1}\bar{q}_{2}$ forming a meson $M_{j}$. Two probabilities
can be further decomposed as
\begin{align}
\frac{\kappa_{B_{j}}}{A_{B_{j}}} & \equiv P_{q_{1}q_{2}q_{3}\rightarrow B_{j}}=C_{B_{j}}N_{iter}\frac{\overline{N}_{B}}{N_{q}^{3}},\label{eq:PBj}\\
	\frac{\kappa_{M_{j}}}{A_{M_{j}}} & \equiv P_{q_{1}\bar{q}_{2}\rightarrow M_{j}}=C_{M_{j}}\frac{\overline{N}_{M}}{N_{q}N_{\bar{q}}}.\label{eq:PMj}
\end{align}
Taking meson for example, $\overline{N}_{M}/N_{q}N_{\bar{q}}$ is
used to denote the averaged probability of a $q\bar{q}$ pair forming
a meson. Here, $N_{q}N_{\bar{q}}$ is the all possible number of $q\bar{q}$
pair, where $N_{q}=N_{u}+N_{d}+N_{s}$ is the number of all quarks
and $N_{\bar{q}}=N_{\bar{u}}+N_{\bar{d}}+N_{\bar{s}}$ is that of
all antiquarks. $\overline{N}_{M}$ is the average number of mesons
produced by the hadronization of quark system with given quark number
$N_{q}$ and antiquark number $N_{\bar{q}}$. The factor $C_{M_{j}}$
denotes further sophisticated tune for the production probability
of $M_{j}$ on the basis of the averaged probability $\overline{N}_{M}/N_{q}N_{\bar{q}}$.
The baryon formula is similar to meson except a factor $N_{iter}$.
This factor is to assure $\sum_{q_{1}q_{2}q_{3}}N_{iter}N_{q_{1}}N_{q_{2}}N_{q_{3}}=N_{q}^{3}$
and equals to $1,3,6$ for $q_{1}q_{2}q_{3}$ with three identical
flavor, two identical flavor, and three different flavors, respectively.
Finally, we obtain yield formulas of hadrons 
\begin{align}
N_{B_{j}} & =C_{B_{j}}N_{iter}N_{q_{1}}N_{q_{2}}N_{q_{3}}\frac{\overline{N}_{B}}{N_{q}^{3}},\label{eq:Nbi}\\
	N_{M_{j}} & =C_{M_{j}}N_{q_{1}}N_{\bar{q}_{2}}\frac{\overline{N}_{M}}{N_{q}N_{\bar{q}}}.\label{eq:Nmi}
\end{align}
 $\overline{N}_{B}$ and $\overline{N}_{M}$ are functions of $N_{q}$
and $N_{\bar{q}}$ \cite{Song:2013isa},

\begin{align}
\overline{N}_{M} & =\frac{x}{2}\left[1-z\frac{\left(1+z\right)^{a}+\left(1-z\right)^{a}}{\left(1+z\right)^{a}-\left(1-z\right)^{a}}\right],\\
\overline{N}_{B} & =\frac{xz}{3}\frac{\left(1+z\right)^{a}}{\left(1+z\right)^{a}-\left(1-z\right)^{a}},\\
\overline{N}_{\bar{B}} & =\frac{xz}{3}\frac{\left(1-z\right)^{a}}{\left(1+z\right)^{a}-\left(1-z\right)^{a}},\label{eq:NBbar}
\end{align}
where $x=N_{q}+N_{\bar{q}}$ and $z=\left(N_{q}-N_{\bar{q}}\right)/x$.
Parameter $a=1+\left(\overline{N}_{M}/\overline{N}_{B}\right)_{z=0}/3$
denotes the production competition of baryon to meson and is tuned
to be $a\approx4.86\pm0.1$ according to our recent work \cite{Shao:2017eok}. 

In this paper, we only consider the production of the ground state
$J^{P}=0^{-},\,1^{-}$ mesons and $J^{P}=(1/2)^{+},\,(3/2)^{+}$ baryons
in flavor SU(3) group. In meson formation, we introduce a parameter
$R_{V/P}$ to describe the relative weight of a quark-antiquark pair
forming the state of spin 1 to that of spin 0. Here, we take $R_{V/P}=0.55\pm0.05$
in order to reproduce the measured $K^{*}/K$ and $\phi/K$ data in
high energy reactions \cite{Adam:2016bpr,Acharya:2019bli}. Factor
$C_{M_{j}}$ is then parameterized as 
\begin{equation}
C_{M_{j}}=\left\{ \begin{array}{ll}
\frac{1}{1+R_{V/P}} & \text{for }J^{P}=0^{-}\textrm{ mesons}\\
\frac{R_{V/P}}{1+R_{V/P}} & \textrm{for }J^{P}=1^{-}\textrm{ mesons}.
\end{array}\right.
\end{equation}
In baryon formation, we introduce a parameter $R_{D/O}$ to describe
the relative weight of spin $3/2$ state to $1/2$ state for three
quark combination. We take $R_{D/O}=0.5\pm0.04$ by fitting the experimental
data of $\Xi^{*}/\Xi$ and $\Sigma^{*}/\Lambda$ in high energy collisions
\cite{Adamova:2017elh}. Then we have for $C_{B_{j}}$
\begin{equation}
C_{B_{j}}=\left\{ \begin{array}{ll}
\frac{1}{1+R_{D/O}} & \textrm{for }J^{P}=({1}/{2})^{+}\textrm{ baryons}\\
\frac{R_{D/O}}{1+R_{D/O}} & \textrm{for }J^{P}=({3}/{2})^{+}\textrm{ baryons},
\end{array}\right.
\end{equation}
except that $C_{\Lambda}=C_{\Sigma^{0}}={1}/{(2+R_{D/O})},~C_{\Sigma^{*0}}={R_{D/O}}/{(2+R_{D/O})},~C_{\Delta^{++}}=C_{\Delta^{-}}=C_{\Omega^{-}}=1$. 

Taking $f_{q_{i}}\left(p\right)$ as model inputs, we can calculate
$f_{h}\left(p\right)$ and $N_{h}$ of hadrons directly produced at
hadronization. Finally, we take the decay contribution of short-life
resonances into account according to experimental measurements, and
obtain results of final-state hadrons
\begin{equation}
f_{h_{j}}^{\left(final\right)}\left(p\right)=f_{h_{j}}\left(p\right)+\sum_{i\neq j}\int dp'f_{h_{i}}\left(p'\right)D_{ij}\left(p',p\right),
\end{equation}
where the decay function $D_{ij}\left(p',p\right)$ is determined
by the decay kinematics and decay branch ratios \cite{Agashe:2014kda}. 

\section{Strangeness neutralization in heavy-ion collisions\label{sec:Sconservation} }

In relativistic heavy-ion collisions, strange quark and antiquark
are always created in pair in collisions and therefore strangeness
is globally conserved. However, for a finite kinetic region such as
the mid-rapidity region, the strangeness neutralization is not so
explicit, in particular, at low collision energies. In this section,
using the precise data for yield densities and $p_{T}$ spectra of
identified hadrons \cite{Adler:2003cb,Adams:2006ke,Abelev:2008aa,Abelev:2008ab,Aggarwal:2010ig,Adamczyk:2017iwn,Adam:2019koz},
we study the local strangeness in the midrapidity region in Au+Au
collisions at STAR BES energies. 

\subsection{strangeness at mid-rapidity }

In this subsection, we estimate strangeness density $dN_{\bar{s}}/dy-dN_{s}/dy$
in the mid-rapidity region in relativistic heavy-ion collisions. We
write $N_{\bar{s}}-N_{s}$ for short in the following. Since experimental
measurements are mainly for hadrons in ground state in flavor SU(3),
we estimate the strangeness by yield densities of the following hadrons
\begin{align}
 & N_{\bar{s}}-N_{s}\nonumber \\
 & =\left(K^{+}+K^{0}+K^{*+}+K^{*0}\right)\nonumber \\
 & -\left(K^{-}+\bar{K}^{0}+K^{*-}+\bar{K}^{*0}\right)\\
 & -\left(\Lambda+\Sigma^{0,\pm}+\Sigma^{*0,*\pm}+2\Xi^{-,0}+2\Xi^{*-,0}+3\Omega^{-}\right)\nonumber \\
 & +\left(\bar{\Lambda}+\bar{\Sigma}^{0,\mp}+\bar{\Sigma}^{*0,*\mp}+2\bar{\Xi}^{+,0}+2\bar{\Xi}^{*+,0}+3\bar{\Omega}^{+}\right).\nonumber 
\end{align}
For convenience, we use $h$ to denote $dN_{h}/dy$. 
The contribution of baryons with different charge states is also abbreviated, e.g., $\Sigma^{0,\pm}\equiv\Sigma^{0}+\Sigma^{-}+\Sigma^{+}$. 
We note that the contribution of higher mass resonances can be effectively included 
if we identify above ground state hadrons as the measured ones. 

The strangeness in meson sector can be calculated as 
\begin{align}
 & K^{+}+K^{0}+K^{*+}+K^{*0}-K^{-}-\bar{K}^{0}-K^{*-}-\bar{K}^{*0}\nonumber \\
 & =\left(K^{+}-K^{-}\right)_{final}+\left(K^{0}-\bar{K}^{0}\right)_{final},
\end{align}
where we use the subscript $final$ to denote that $K^{+}-K^{-}$
and $K^{0}-\bar{K}^{0}$ have received the decay contribution of $K^{*}$.
Since neutral kaons are not measured, we use the approximation 
\begin{equation}
\left(K^{+}-K^{-}\right)_{final}\approx\left(K^{0}-\bar{K}^{0}\right)_{final}.\label{eq:net_kaon}
\end{equation}

For strangeness contained in hyperons with one strange quark, we decompose
them into two parts: experimentally measured net-$\Lambda$ and experimentally
un-measured net-$\Sigma^{\pm}$. Using the property of S\&EM decays
for hyperons \cite{Agashe:2014kda}, we have 
\begin{align}
 & \left(\Lambda+\Sigma^{0,\pm}+\Sigma^{*0,*\pm}\right)-\left(\bar{\Lambda}+\bar{\Sigma}^{0,\mp}+\bar{\Sigma}^{*0,*\mp}\right)\nonumber \\
 & =\left(\Lambda-\bar{\Lambda}\right)_{final}+\left(\Sigma^{\pm}-\bar{\Sigma}^{\mp}\right)_{final}
\end{align}
with
\begin{align}
 & \left(\Lambda-\bar{\Lambda}\right)_{final}\\
 & =\left(\Lambda-\bar{\Lambda}\right)+\left(\Sigma^{0}-\bar{\Sigma}^{0}\right)+0.94\left(\Sigma^{*\pm}-\bar{\Sigma}^{*\mp}\right)\nonumber \\
 & +0.88\left(\Sigma^{*0}-\bar{\Sigma}^{*0}\right)\nonumber 
\end{align}
and 
\begin{align}
 & \left(\Sigma^{\pm}-\bar{\Sigma}^{\mp}\right)_{final}\\
 & =\left(\Sigma^{\pm}-\bar{\Sigma}^{\mp}\right)+0.06\left(\Sigma^{*\pm}-\bar{\Sigma}^{*\mp}\right)+0.12\left(\Sigma^{*0}-\bar{\Sigma}^{*0}\right).\nonumber 
\end{align}
Applying our formula of hadronic yields in Sec. \ref{sec:sdqcm},
we obtain 
\begin{equation}
\frac{\left(\Sigma^{\pm}-\bar{\Sigma}^{\mp}\right)_{final}}{\left(\Lambda-\bar{\Lambda}\right)_{final}}=\frac{1.1+0.68R_{D/O}+0.099R_{D/O}^{2}}{1.096+2.62R_{D/O}+R_{D/O}^{2}}\approx0.55.
\end{equation}

For strangeness contained in hyperons with two strange quarks, after
considering S\&EM decays, we have 
\begin{equation}
\left(\Xi^{-,0}-\bar{\Xi}^{+,0}\right)+\left(\Xi^{*-,0}-\bar{\Xi}^{*+,0}\right)=\left(\Xi^{-,0}-\bar{\Xi}^{+,0}\right)_{final}
\end{equation}
and we use the approximation 
\begin{equation}
\left(\Xi^{-}-\bar{\Xi}^{+}\right)_{final}\approx\left(\Xi^{0}-\bar{\Xi}^{0}\right)_{final}.\label{eq:net_Xi}
\end{equation}

By the sum over the strangeness in meson and baryon sectors, we obtain
the net-strangeness of the system 
\begin{align}
N_{\bar{s}}-N_{s} & \approx2\left(K^{+}-K^{-}\right)_{final}-1.57\left(\Lambda-\bar{\Lambda}\right)_{final}\nonumber \\
 & -4\left(\Xi^{-}-\bar{\Xi}^{+}\right)_{final}-3\left(\Omega^{-}-\bar{\Omega}^{+}\right),\label{eq:net_S}
\end{align}
where subscript $final$ denotes the yield including S\&EM decay contributions. 

The total number of strange quarks and strange anti-quarks is  
\begin{align}
N_{\bar{s}}+N_{s} & \approx2\left(K^{+}+K^{-}\right)_{final}+1.57\left(\Lambda+\bar{\Lambda}\right)_{final}\nonumber \\
 & +4\left(\Xi^{-}+\bar{\Xi}^{+}\right)_{final}+3\left(\Omega^{-}+\bar{\Omega}^{+}\right)\nonumber \\
 & +2\text{\ensuremath{\phi}}+2\left(\frac{2}{3}\ensuremath{\eta}+\frac{1}{3}\ensuremath{\eta'}\right).\label{eq:tot_S}
\end{align}

Because the strangeness $N_{\bar{s}}-N_{s}$ is explicitly dependent
on collision energies and collision centralities, we define the relative
asymmetry factor 
\begin{equation}
z_{S}=\frac{N_{\bar{s}}-N_{s}}{N_{\bar{s}}+N_{s}},
\end{equation}
which is convenient to compare results in different situations.

\begin{table*}[!htb]
\caption{Strangeness asymmetry factor $z_{S}$ calculated by yield data of
strange hadrons and anti-hadrons in central Au+Au collisions \cite{Adler:2003cb,Adams:2006ke,Abelev:2008aa,Abelev:2008ab,Aggarwal:2010ig,Adamczyk:2015lvo,Adam:2019koz}.} \label{tab:zs}%
\begin{tabular}{ccccccc}
\toprule 
$\sqrt{s_{NN}}$ (GeV) & $K^{+,-}$ & $\Lambda\left(\bar{\Lambda}\right)$ & $\Xi^{-}\left(\bar{\Xi}^{+}\right)$ & $\Omega^{-}\left(\bar{\Omega}^{+}\right)$ & $\phi$ & $z_{S}$\tabularnewline
\midrule
\midrule 
200 & $\begin{array}{c}
48.9\pm6.3\\
45.7\pm5.2
\end{array}$ & $\begin{array}{c}
16.7\pm1.1\\
12.7\pm0.9
\end{array}$ & $\begin{array}{c}
2.17\pm0.2\\
1.83\pm0.2
\end{array}$ & $0.53\pm0.06$ & $7.95\pm0.11$ & $-0.004\pm0.06$\tabularnewline
\midrule 
62.4 & $\begin{array}{c}
37.6\pm2.7\\
32.4\pm2.3
\end{array}$ & $\begin{array}{c}
15.7\pm2.3\\
8.3\pm1.1
\end{array}$ & $\begin{array}{c}
1.63\pm0.2\\
1.03\pm0.11
\end{array}$ & $\begin{array}{c}
0.212\pm0.028\\
0.167\pm0.027
\end{array}$ & $3.52\pm0.08$ & $-0.019\pm0.04$\tabularnewline
\midrule 
39 & $\begin{array}{c}
32.0\pm2.9\\
25.0\pm2.3
\end{array}$ & $\begin{array}{c}
11.02\pm0.03\\
3.82\pm0.01
\end{array}$ & $\begin{array}{c}
1.54\pm0.01\\
0.78\pm0.01
\end{array}$ & $\begin{array}{c}
0.191\pm0.006\\
0.139\pm0.004
\end{array}$ & $3.38\pm0.03$ & $-0.002\pm0.05$\tabularnewline
\midrule 
27 & $\begin{array}{c}
31.1\pm2.8\\
22.6\pm2.0
\end{array}$ & $\begin{array}{c}
11.67\pm0.04\\
2.75\pm0.01
\end{array}$ & $\begin{array}{c}
1.57\pm0.01\\
0.598\pm0.006
\end{array}$ & $\begin{array}{c}
0.154\pm0.008\\
0.0972\pm0.0049
\end{array}$ & $3.01\pm0.04$ & $-0.006\pm0.05$\tabularnewline
\midrule 
19.6 & $\begin{array}{c}
29.6\pm2.9\\
18.8\pm1.9
\end{array}$ & $\begin{array}{c}
12.58\pm0.04\\
1.858\pm0.009
\end{array}$ & $\begin{array}{c}
1.62\pm0.02\\
0.421\pm0.005
\end{array}$ & $\begin{array}{c}
0.155\pm0.01\\
0.0811\pm0.0048
\end{array}$ & $2.57\pm0.04$ & $-0.0002\pm0.05$\tabularnewline
\midrule 
11.5 & $\begin{array}{c}
25.0\pm2.5\\
12.3\pm1.2
\end{array}$ & $\begin{array}{c}
14.17\pm0.08\\
0.659\pm0.009
\end{array}$ & $\begin{array}{c}
1.35\pm0.02\\
0.169\pm0.004
\end{array}$ & $\begin{array}{c}
0.082\pm0.012\\
0.0356\pm0.0052
\end{array}$ & $1.72\pm0.04$ & $-0.004\pm0.05$\tabularnewline
\midrule 
7.7 & $\begin{array}{c}
20.8\pm1.7\\
7.7\pm0.6
\end{array}$ & $\begin{array}{c}
15.3\pm0.11\\
0.193\pm0.006
\end{array}$ & $\begin{array}{c}
1.19\pm0.03\\
0.0667\pm0.0044
\end{array}$ & $\begin{array}{c}
0.0271\pm0.0048\\
0.0075\pm0.0013
\end{array}$ & $\begin{array}{c}
1.21\pm0.06\end{array}$ & $-0.021\pm0.04$\tabularnewline
\bottomrule
\end{tabular}
\end{table*}

In Table \ref{tab:zs}, we show results for $z_{S}$ in most central
Au+Au collisions at different collisions energies \footnote{In calculations, we use the approximation of isospin symmetry between up and down quarks in Eqs.~(\ref{eq:net_kaon}) and (\ref{eq:net_Xi}). If we consider the small asymmetry in number between up quarks and down quarks coming from colliding nucleons, we should modify Eq.~(\ref{eq:net_kaon}) by multiplying a factor $N_{u}/N_{d}$ and Eq.~(\ref{eq:net_Xi}) by multiplying a factor $(N_{u}/N_{d})^{-1}$ in right hand side of equation and the corresponding coefficients in Eqs. (\ref{eq:net_S}) and (\ref{eq:tot_S}).  According to the number of newborn quarks extracted in next section and net-quarks from participant nucleons, we obtain, for example, $N_{u}/N_{d}\approx0.99$ at $\sqrt{s_{NN}}=$ 200 GeV and $N_{u}/N_{d}\approx0.93$ at 7.7 GeV. The resulting  $z_{S}$ are (-0.004, -0.018, -0.0014, -0.005, 0.002, 0.003, -0.009) at $\sqrt{s_{NN}}=$ (200, 62.4, 39, 27, 19.6, 11.5, 7.7) GeV, respectively. They are very close to those shown in Tab.~\ref{tab:zs}.} Experimental data for yield densities of hadrons that are used in
calculation are also presented \cite{Adler:2003cb,Adams:2006ke,Abelev:2008aa}
\cite{Abelev:2008ab,Aggarwal:2010ig,Abelev:2008aa} \cite{Adamczyk:2015lvo,Adam:2019koz}.
Because some data for $\Omega^{-}$ and $\phi$ are results in 0-10\%
centrality, we re-scale them by multiplying a factor $N_{part}^{(0-5\%)}/N_{part}^{(0-10\%)}$
according to the participant nucleon number $N_{part}$. We do the
similar re-scaling for data of $\Omega^{-}$ in 0-60\% centrality
at $\sqrt{s_{NN}}=$ 7.7 GeV. Because data of $\eta$ and $\eta'$
mesons are usually absent, we neglect them and therefore the calculated
$z_{S}$ is over-estimated. We see that $z_{S}$ at seven collision
energies are quite small. 
Therefore, strangeness neutralization $N_{s}=N_{\bar{s}}$ is well satisfied in mid-rapidity region in heavy-ion collisions at RHIC energies. 

\subsection{$p_{T}$ spectrum symmetry}

In this subsection, we study the symmetry property between $p_{T}$
spectrum of strange quark $f_{s}\left(p_{T}\right)$ and that of strange
antiquark $f_{\bar{s}}\left(p_{T}\right)$. For this purpose, we choose
$\Omega^{-}$ ($\bar{\Omega}^{+}$) which consists of only strange
quarks (antiquarks). Using Eq. (\ref{eq:fbfinal}), we have 
\begin{align}
f_{\Omega}\left(3p_{T}\right) & =\kappa_{\Omega}f_{s}^{3}\left(p_{T}\right),\\
f_{\bar{\Omega}}\left(3p_{T}\right) & =\kappa_{\bar{\Omega}}f_{\bar{s}}^{3}\left(p_{T}\right)
\end{align}
from which we have 
\begin{equation}
\frac{f_{\bar{s}}\left(p_{T}\right)}{f_{s}\left(p_{T}\right)}=\kappa_{\bar{\Omega},\Omega}\left[\frac{f_{\bar{\Omega}}\left(3p_{T}\right)}{f_{\Omega}\left(3p_{T}\right)}\right]^{1/3}\propto\left[\frac{f_{\bar{\Omega}}\left(3p_{T}\right)}{f_{\Omega}\left(3p_{T}\right)}\right]^{1/3},\label{eq:rssbar}
\end{equation}
where $\kappa_{\bar{\Omega},\Omega}=\left(\kappa_{\bar{\Omega}}/\kappa_{\Omega}\right)^{1/3}$
is independent of $p_{T}$ but is dependent on quark numbers. We emphasize
that $\kappa_{\bar{\Omega},\Omega}$ is not equal to one at nonzero
net quark number. 

\begin{figure}[h]
\includegraphics[width=0.95\linewidth]{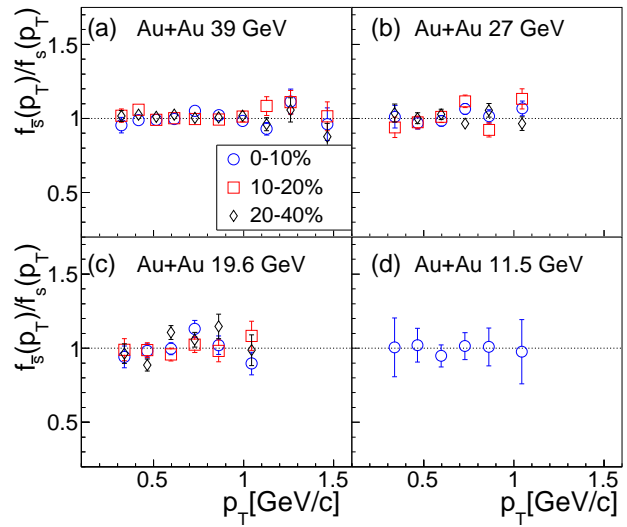}
    \caption{Ratio $f_{\bar{s}}\left(p_{T}\right)/f_{s}\left(p_{T}\right)$ in
Au+Au collisions at different collision energies obtained from experimental
data of $\Omega^{-}$ and $\bar{\Omega}^{+}$ in central and semi-central
collisions \cite{Adamczyk:2015lvo,Adam:2019koz} by Eq. (\ref{eq:rssbar}).
\label{fig:ssbarRatio}}

\end{figure}

Using data of $p_{T}$ spectra for $\Omega^{-}$ and $\bar{\Omega}^{+}$
in central Au+Au collisions \cite{Adamczyk:2015lvo,Adam:2019koz},
we calculate the ratio $f_{\bar{s}}\left(p_{T}\right)/f_{s}\left(p_{T}\right)$
at different collision energies and present results in Fig. \ref{fig:ssbarRatio}.
Since we have shown $N_{s}=N_{\bar{s}}$ in the previous subsection,
we multiply a proper constant before data of $f_{\bar{\Omega}}^{1/3}\left(3p_{T}\right)/f_{\Omega}^{1/3}\left(3p_{T}\right)$
to satisfy $N_{s}=N_{\bar{s}}$ and therefore we can directly compare
$f_{\bar{s}}\left(p_{T}\right)/f_{s}\left(p_{T}\right)$ in/at different
collision centralities/energies. Because of finite statistics of $\Omega^{-}$
and $\bar{\Omega}^{+}$, data points of $f_{\bar{s}}\left(p_{T}\right)/f_{s}\left(p_{T}\right)$
show a certain fluctuations around one. On the whole, we can see that
$f_{\bar{s}}\left(p_{T}\right)=f_{s}\left(p_{T}\right)$ is a good
approximation at the studied collision energies. 

\section{$p_{T}$ spectra of hadrons \label{sec:hpt}}

In this section, we use the quark combination model (QCM) in Sec.  \ref{sec:sdqcm} to study $p_{T}$ spectra of light-flavor hadrons in Au+Au collisions at RHIC energies. The inputs of model are $p_{T}$ spectra of quarks and antiquarks at hadronization. Here, we take $f_{\bar{s}}\left(p_{T}\right)=f_{s}\left(p_{T}\right)$ based on the study of strangeness neutralization in Sec. \ref{sec:Sconservation}.
 We take $f_{\bar{u}}\left(p_{T}\right)=f_{\bar{d}}\left(p_{T}\right)$ for the newborn up and down anti-quarks.  Because a part of up and down quarks comes from the colliding nucleons, $p_T$ spectrum of up quarks is not exactly the same as that of down quarks.  As discussed in [47], the relative difference in number between up and down quarks is only a few percentages.  We have checked that $p_T$ spectra of hadrons and yield ratios of anti-hadron to hadron studied in this paper are not sensitive to this small asymmetry.  Therefore, in this paper, we take the approximation $f_{u}\left(p_{T}\right)\approx f_{d}\left(p_{T}\right)$ in the mid-rapidity region.

\begin{figure}
\includegraphics[width=0.95\linewidth]{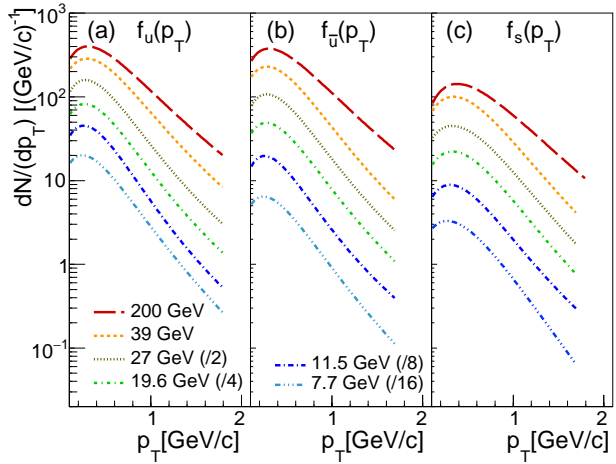}
    \caption{$p_{T}$ spectra of quarks at hadronization at mid-rapidity in 
Au+Au collisions for 0-5\% centrality. Spectra at $\sqrt{s_{NN}}=$ 7.7-27 GeV are scaled
for clarity as shown in the figure.}
\label{fig:fuds}
\end{figure}

In Table \ref{tab:inpVvar}, we list all inputs and parameters of the model which are needed to calculate $p_T$ spectra of hadrons.  As introduced in model description in Sec. \ref{sec:sdqcm}, two parameters $R_{V/P}$ and $R_{D/O}$ are taken as 0.55 and 0.5, respectively, at all the studied collision energies. 
For three inputs, $f_{s}\left(p_{T}\right)$ is fixed by using our model to fit experimental data of $\phi$, and $f_{u}\left(p_{T}\right)$ ($f_{\bar{u}}\left(p_{T}\right)$) is fixed by experimental data 
of (anti-)baryons containing $u(\bar{u})$ such as (anti-)proton \cite{Abelev:2008ab,Adamczyk:2017iwn,Adamczyk:2015lvo}, respectively. The extracted results for quark $p_{T}$ spectra in Au+Au collisions for 0-5\% centrality at six RHIC energies are shown in Fig. \ref{fig:fuds}. 

\begin{table}
\caption{Inputs and parameters of the model to calculation $p_T$ spectra of hadrons.}
\label{tab:inpVvar}%
    \begin{tabular*}{7cm}{@{\extracolsep{\fill}}ccccc}
\toprule 
    \multicolumn{3}{c}{input} & \multicolumn{2}{c}{parameter} \tabularnewline
\midrule 
     $f_{u}(p_T)$ & $f_{\bar{u}}(p_T)$ & $f_{s}(p_T)$  & $R_{V/P}$  & $R_{O/D}$  \tabularnewline
\bottomrule
\end{tabular*}
\end{table}

\begin{figure}
\includegraphics[width=0.95\linewidth]{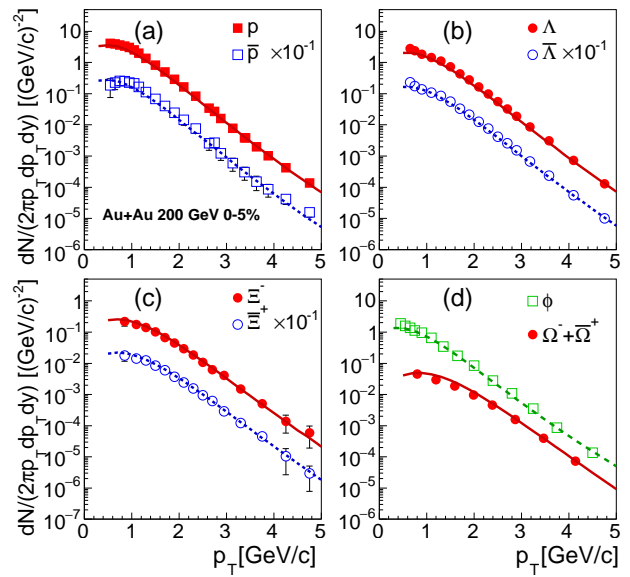}
    \caption{$p_{T}$ spectra of hadrons at mid-rapidity in central Au+Au collisions
at $\sqrt{s_{NN}}=$ 200 GeV. Symbols are experimental data \cite{Abelev:2006jr,Adams:2006ke,Abelev:2008aa}
and lines are results of our model. Spectra of some hadrons are scaled
for clarity as shown in the figure.}

\label{fig:hpt200GeV}
\end{figure}
\begin{figure}
\includegraphics[width=0.95\linewidth]{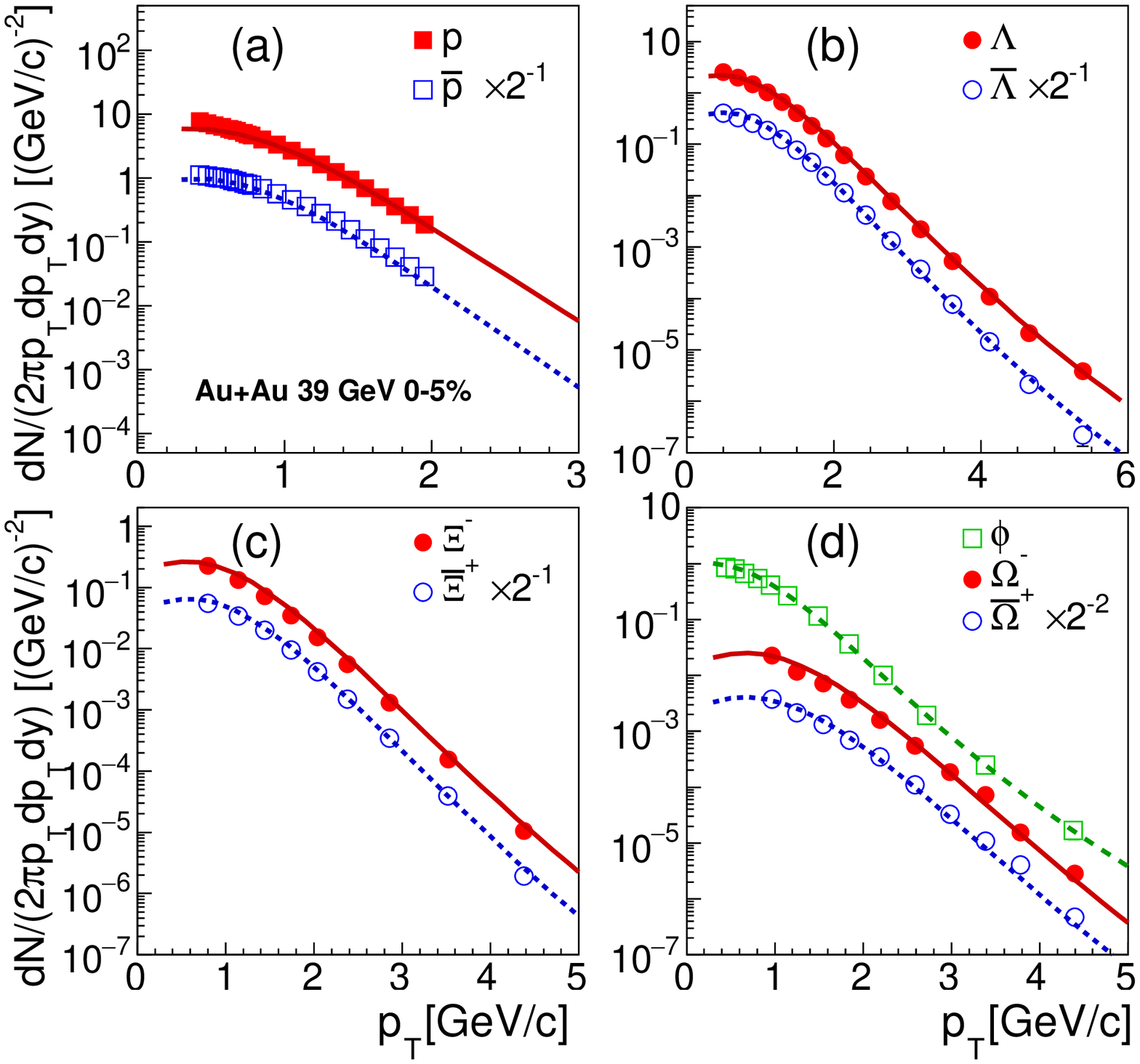}
	\caption{Same as Fig. \ref{fig:hpt200GeV} but for central Au+Au collisions
at $\sqrt{s_{NN}}=39$ GeV. Experimental data are from \cite{Adamczyk:2015lvo,Adamczyk:2017iwn,Adam:2019koz}. }

\label{fig:hpt39GeV}
\end{figure}
\begin{figure}
\includegraphics[width=0.95\linewidth]{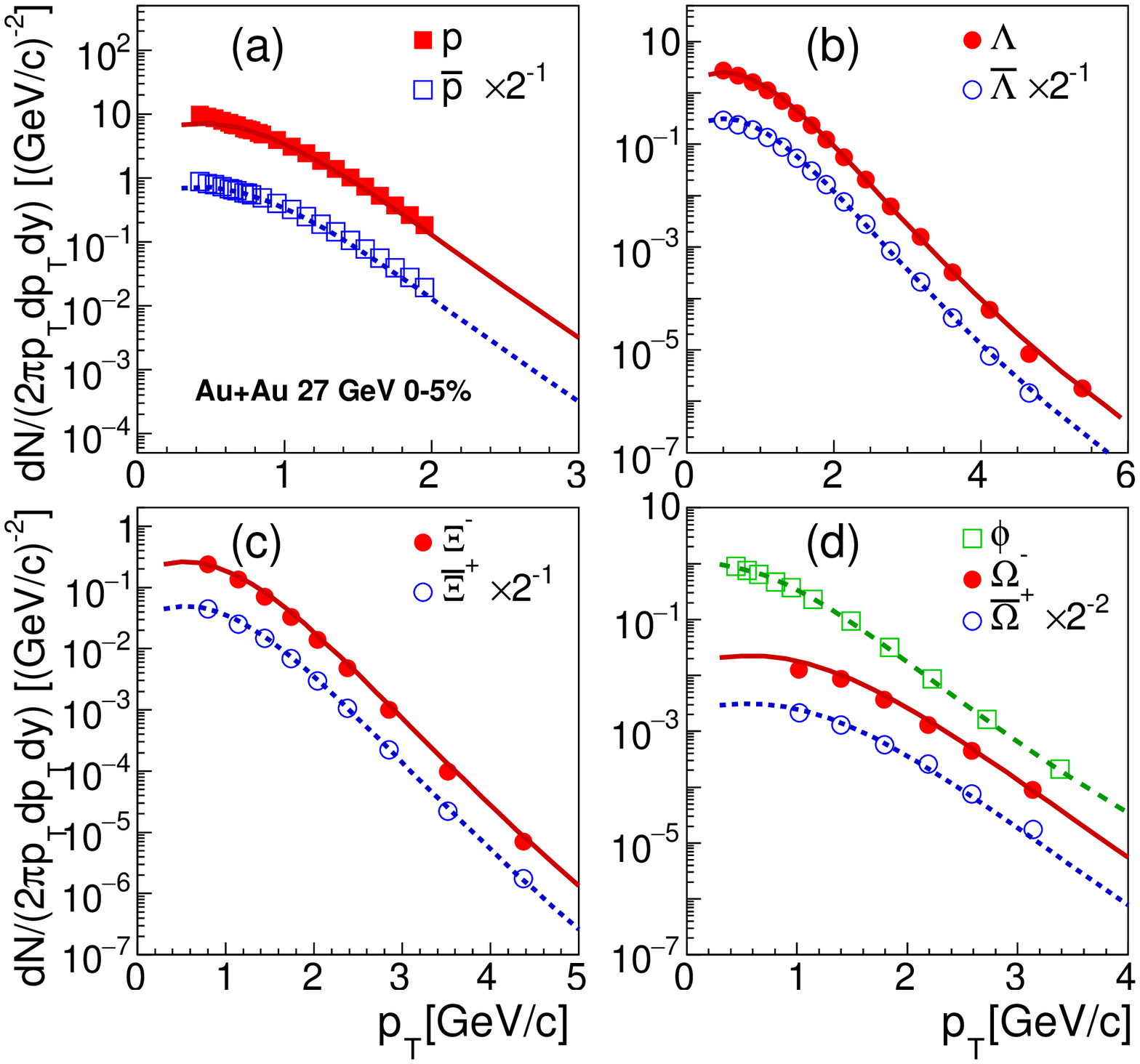}
	\caption{Same as Fig. \ref{fig:hpt200GeV} but for central Au+Au collisions
at $\sqrt{s_{NN}}=27$ GeV. Experimental data are from \cite{Adamczyk:2015lvo,Adamczyk:2017iwn,Adam:2019koz}. }
\label{fig:hpt27GeV}
\end{figure}
\begin{figure}
\includegraphics[width=0.95\linewidth]{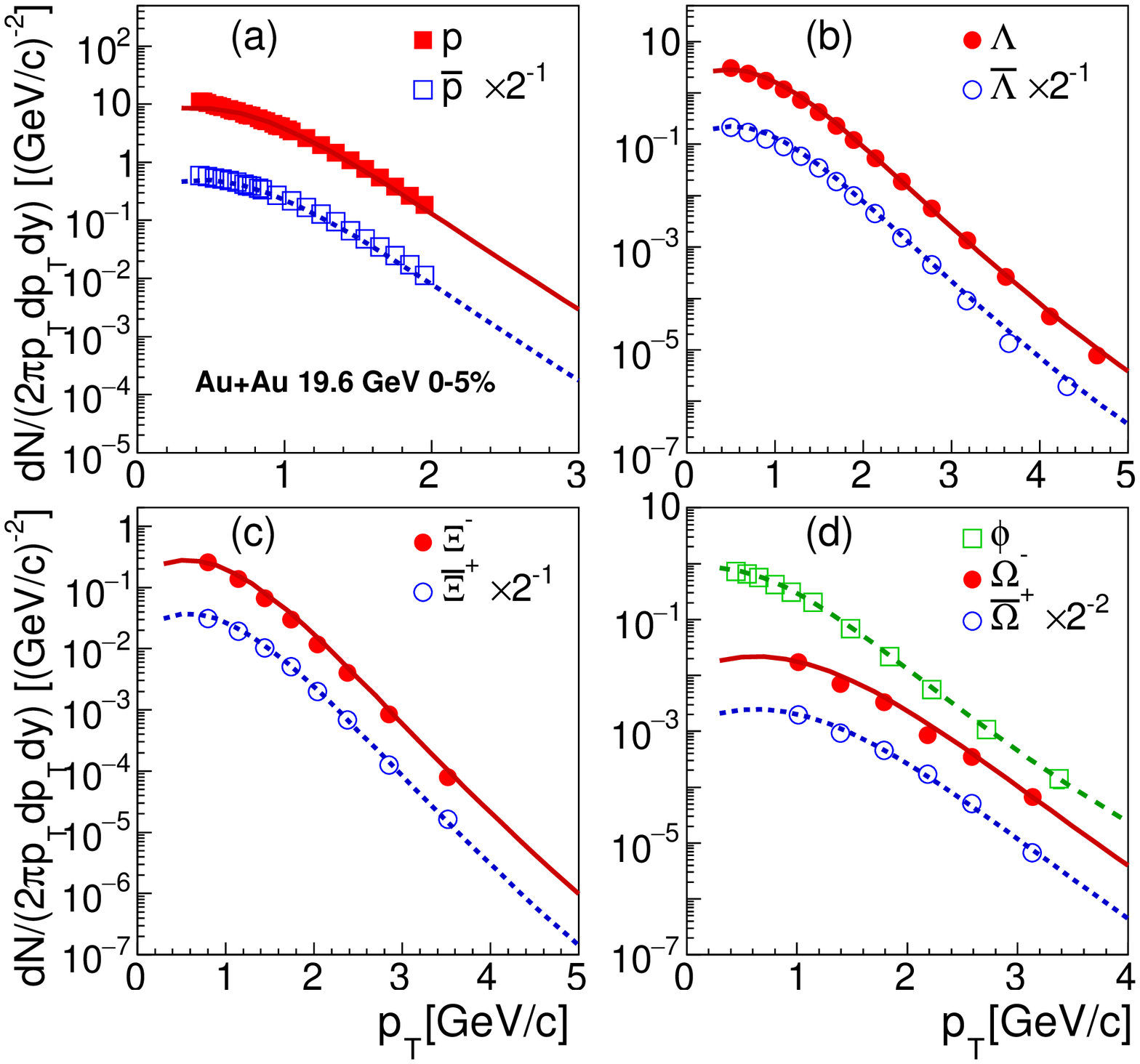}
	\caption{Same as Fig. \ref{fig:hpt200GeV} but for central Au+Au collisions at $\sqrt{s_{NN}}=19.6$ GeV. Experimental data are from \cite{Adamczyk:2015lvo,Adamczyk:2017iwn,Adam:2019koz}. }
\label{fig:hpt19GeV}
\end{figure}
\begin{figure}
\includegraphics[width=0.95\linewidth]{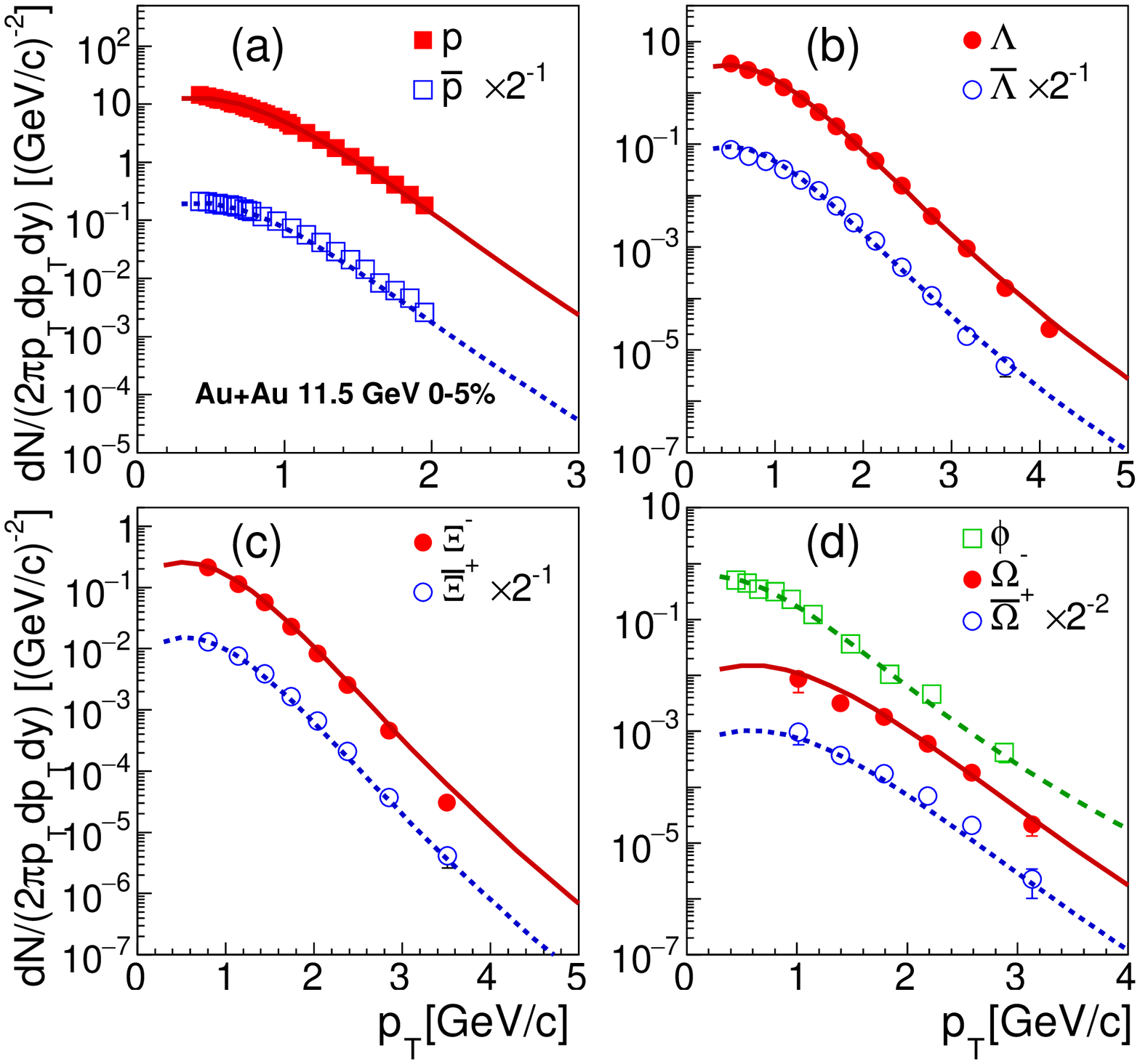}
	\caption{Same as Fig. \ref{fig:hpt200GeV} but for central Au+Au collisions at $\sqrt{s_{NN}}=11.5$ GeV. Experimental data are from \cite{Adamczyk:2015lvo,Adamczyk:2017iwn,Adam:2019koz}. }
\label{fig:hpt11GeV}
\end{figure}

\begin{figure}
\includegraphics[scale=0.4]{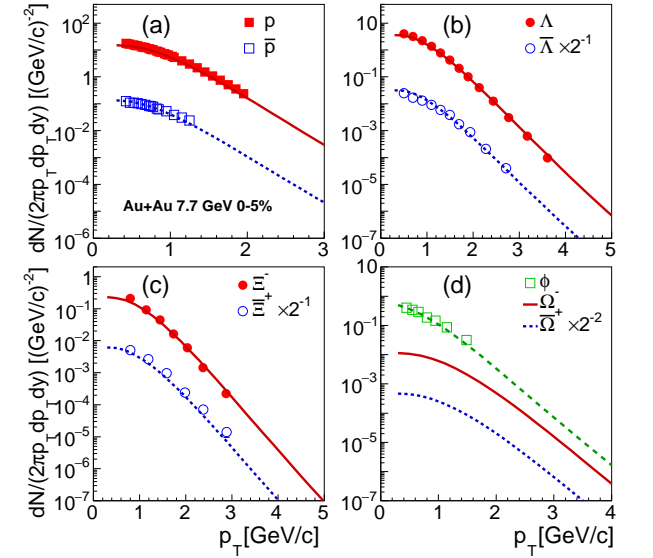}\caption{Same as Fig. \ref{fig:hpt200GeV} but for central Au+Au collisions
at $\sqrt{s_{NN}}=$ 7.7 GeV. Experimental data are from \cite{Adamczyk:2015lvo,Adamczyk:2017iwn,Adam:2019koz}. }
\label{fig:hpt7GeV}
\end{figure}

In Fig. \ref{fig:hpt200GeV}, we show the calculated results for $p_{T}$
spectra of hadrons in central Au+Au collisions at $\sqrt{s_{NN}}=$
200 GeV and compare them with experimental data \cite{Abelev:2006jr,Adams:2006ke,Abelev:2008aa}.
We see that the agreement between our model results and experimental
data is satisfactory. Although there exists many successful explanations
on these experimental data in the framework of quark combination mechanism
in previous works \cite{Fries:2003vb,Greco:2003xt,Hwa:2002tu,Chen:2006vc,Shao:2004cn,Shao:2009uk},
here we would like to emphasize that the current EVC model can systematically
explain these data in a simple way. Furthermore, the good agreement
at top RHIC energy provides important basis for the application of
our model to lower RHIC energies. 

In Figs. \ref{fig:hpt39GeV}, \ref{fig:hpt27GeV}, \ref{fig:hpt19GeV},
\ref{fig:hpt11GeV} and \ref{fig:hpt7GeV}, we show results for $p_{T}$
spectra of hadrons in central Au+Au collisions at $\sqrt{s_{NN}}=$
39, 27, 19.6, 11.5 and 7.7 GeV and compare with experimental data
\cite{Adamczyk:2015lvo,Adamczyk:2017iwn,Adam:2019koz}. At $\sqrt{s_{NN}}=$
39, 27, 19.6, 11.5 GeV, we see a good agreement between our model
results and experimental data \cite{Adamczyk:2015lvo,Adamczyk:2017iwn,Adam:2019koz}.
In particular, we see that experimental data for baryons ($p$, $\Lambda$,
$\Xi$, $\Omega$) and meson $\phi$ can be explained by the model
very well. At $\sqrt{s_{NN}}=$ 7.7 GeV, we also see that model results
are in good agreement with available experimental data. However, compared
with data in Figs. \ref{fig:hpt200GeV}-\ref{fig:hpt11GeV}, the available
data at $\sqrt{s_{NN}}=$ 7.7 GeV cover smaller $p_{T}$ range and
$\Omega$ data in the most central collisions are absent. Therefore
the comparison at $\sqrt{s_{NN}}=$ 7.7 GeV is not as conclusive as
those at other five higher energies. We should study this energy point
further in the future when more precise data are available. 

Using these systematic calculations and comparisons, we would like
to emphasize the equal-velocity combination of quarks and antiquarks
as the effective description at hadronization. This is manifested
by the following two points. First, from panel (d) in Figs. \ref{fig:hpt200GeV}-\ref{fig:hpt11GeV},
we see that experimental data of $\Omega$ and $\phi$ can be perfectly
explained by the same $f_{s}\left(p_{T}\right)$ in a very simple
way, 
\begin{align}
f_{\Omega}\left(p_{T}\right) & =\kappa_{\Omega}f_{s}^{3}\left(p_{T}/3\right),\\
f_{\phi}\left(p_{T}\right) & =\kappa_{\phi}f_{s}^{2}\left(p_{T}/2\right).
\end{align}
We call this property as the quark number scaling for hadronic $p_{T}$
spectra. We have shown that this property not only exists in relativistic
heavy-ion collisions but also exists in high-multiplicity $pp$ and
$p$-Pb collisions at LHC energies \cite{Song:2017gcz,Zhang:2018vyr,Song:2019sez}.
Second, we see that data of $\Lambda$ and $\Xi^{-}$ can be simultaneously
explained by $f_{s}\left(p_{T}\right)$ from $\phi$ and $f_{u}\left(p_{T}\right)$
from proton, 
\begin{align}
f_{\Lambda}\left(p_{T}\right) & =\kappa_{\Lambda}f_{u}^{2}\left(\frac{1}{2+r}p_{T}\right)f_{s}\left(\frac{r}{2+r}p_{T}\right),\\
f_{\Xi}\left(p_{T}\right) & =\kappa_{\Xi}f_{u}\left(\frac{1}{1+2r}p_{T}\right)f_{s}^{2}\left(\frac{r}{1+2r}p_{T}\right),
\end{align}
after further including the decay contribution of heavier baryons.  Here, $r=m_{s}/m_{u}$. This is a clear support for the equal-velocity combination for quarks with different flavors.


\section{Hadronic yields and multi-particle correlations \label{sec:Yields}}

In this section, we study the $p_{T}$-integrated yields of identified
hadrons. In heavy-ion collisions at RHIC energies, the net baryon
numbers deposited in the mid-rapidity region will influence the production
of hadrons and antihadrons to a certain extent. One of the consequences
for non-zero baryon number density is the asymmetry in yield between
hadrons and their anti-particles. We study this yield asymmetry with
our model by focusing on multi-particle yield correlations. 

\begin{figure}[!htp]
\includegraphics[scale=0.42]{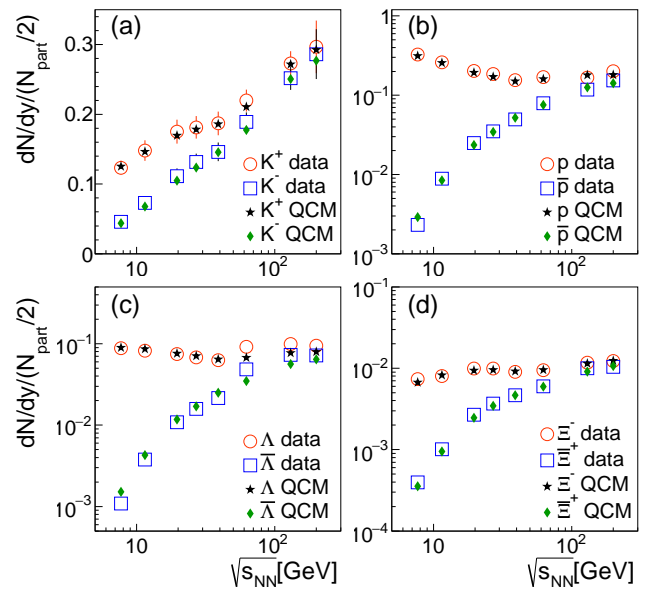}\caption{Hadronic yield densities divided by participant nucleon number at
mid-rapidity in Au+Au collisions at different collision energies.
Open symbols are experimental data \cite{Adams:2003fy,Adams:2006ke,Abelev:2008ab,Aggarwal:2010ig,Adamczyk:2017iwn,Adam:2019koz,Adcox:2002au} and solid symbols are model results.}
\label{fig:dndy}
\end{figure}

In Fig. \ref{fig:dndy}, we show yield densities of hadrons and anti-hadrons \footnote{Model results for yield densities of kaons are presented here. Because the direct combination $u+\bar{s}\to K$ in the current EVC model has energy conservation issue, we have to introduce further treatment to reconcile this such as we did in Ref. \cite{Gou:2017foe}. However, the strange quantum number conservation ensures that the number of the formed kaon can be correctly calculated in the current model.} divided by participant nucleon number at mid-rapidity in central Au+Au collisions at different collision energies.  Open symbols are experimental
data \cite{Adams:2003fy,Adams:2006ke,Abelev:2008ab,Aggarwal:2010ig,Adamczyk:2017iwn,Adam:2019koz,Adcox:2002au}
and solid symbols are model results. Experimental data show that the
yield split between $K^{+}$ and $K^{-}$ in (a) is much smaller than
that between $p$ and $\bar{p}$ in (b). Yield split between $\Lambda$
and $\bar{\Lambda}$ in (c) and that between $\Xi^{-}$ and $\bar{\Xi}^{+}$
in (d) are larger than kaon split in (a) but are smaller than proton
split in (b). Comparing with experimental data, we see that the model
provides a globally good description for yield densities of hadrons
and anti-hadrons. 

In order to further study the split in yield between hadrons and anti-hadrons,
we calculate the ratio of yield for anti-hadron to that for hadron.
Yield ratio is not sensitive to the numbers of quarks and antiquarks
at hadronization and is also not sensitive to some hadronization details
such as parameters $R_{V/P}$ and $R_{D/O}$ introduced in our model.
Therefore, it can be used to provide a more direct test for the quark
combination mechanism at hadronization in relativistic heavy-ion collisions.
Experimentally measured protons and anti-protons contain complex decay
contributions of heavier baryons and anti-baryons, 

\begin{align}
p^{(final)} & =p+\Delta^{++}+\frac{2}{3}\Delta^{+}+\frac{1}{3}\Delta^{0}+0.64\Lambda+0.517\Sigma^{+}\nonumber \\
 & +0.64\Sigma^{0}+0.633\Sigma^{*+}+0.594\Sigma^{*0}+0.602\Sigma^{*-}\nonumber \\
 & +0.64\left(\Xi^{0}+\Xi^{-}+\Xi^{*0}+\Xi^{*-}\right)+0.64\Omega^{-},
\end{align}
where we use the particle name with superscript $\left(final\right)$
to denote the yield density of final-state hadron receiving the decay
contributions and the particle name without superscript in the right
hand side of the equation to denote yields of directly-produced hadrons
by hadronization. The ratio $\bar{p}/p$ finally behaves as 

\begin{equation}
\frac{\bar{p}^{\left(final\right)}}{p^{(final)}}\approx\left(\frac{1-z}{1+z}\right)^{0.99a},\label{eq:rppbar}
\end{equation}
where $z$ is net-quark fraction and the factor $a\approx4.86$ is
related to the baryon-meson production competition \cite{Song:2013isa,Shao:2017eok},
see Eq. (\ref{eq:NBbar}) and texts below. For yields of kaons, we
take the decay contributions of $K^{*}\left(892\right)$ and $\phi$
into account and have 

\begin{align}
\frac{K^{-\left(final\right)}}{K^{+\left(final\right)}} & =\frac{K^{-}+\frac{1}{3}K^{*-}+\frac{2}{3}K^{*0}+0.489\phi}{K^{+}+\frac{1}{3}K^{*+}+\frac{2}{3}K^{*0}+0.489\phi}\nonumber \\
 & =\frac{1-z}{1+z}\frac{1+0.489C_{\phi}\lambda_{s}}{1+\lambda_{s}\left(\frac{z}{1+z}+0.489C_{\phi}\frac{1-z}{1+z}\right)},\label{eq:rKpm}
\end{align}
where $\lambda_{s}=N_{s}/N_{\bar{u}}$. 

For yields of $\Lambda$, $\Xi^{-}$ and their anti-particles, we
consider the S\&EM decay contributions, 
\begin{align}
\frac{\bar{\Lambda}^{\left(final\right)}}{\Lambda^{\left(final\right)}} & =\frac{\bar{\Lambda}+\bar{\Sigma}^{0}+0.94\left(\bar{\Sigma}^{*-}+\bar{\Sigma}^{*+}\right)+0.88\bar{\Sigma}^{*0}}{\Lambda+\Sigma^{0}+0.94\left(\Sigma^{*+}+\Sigma^{*-}\right)+0.88\Sigma^{*0}}\nonumber \\
 & =\left(\frac{1-z}{1+z}\right)^{a-1}\left(1+\lambda_{s}\frac{z}{1+z}\right)^{-2},\label{eq:rLam}
\end{align}
and
\begin{align}
\frac{\bar{\Xi}^{+\left(final\right)}}{\Xi^{-\left(final\right)}} & =\frac{\bar{\Xi}^{+}+\frac{1}{3}\bar{\Xi}^{*+}+\frac{2}{3}\bar{\Xi}^{*0}}{\Xi^{-}+\frac{1}{3}\Xi^{*-}+\frac{2}{3}\Xi^{*0}}\nonumber \\
 & =\left(\frac{1-z}{1+z}\right)^{a-2}\left(1+\lambda_{s}\frac{z}{1+z}\right)^{-1}.\label{eq:rXi}
\end{align}
For $\Omega^{-}$, we directly have 
\begin{equation}
\frac{\bar{\Omega}^{+}}{\Omega^{-}}=\left(\frac{1-z}{1+z}\right)^{a-3}.\label{eq:rOmg}
\end{equation}
In Eqs. (\ref{eq:rKpm}-\ref{eq:rXi}), the power term $\left[\left(1-z\right)/\left(1+z\right)\right]^{\alpha}$
dominates the behavior of three ratios. Therefore, ratios $K^{-}/K^{+}$,
$\bar{p}/p$, $\bar{\Lambda}/\Lambda$, $\bar{\Xi}^{+}/\Xi^{-}$ and
$\bar{\Omega}^{+}/\Omega^{-}$ in our model are simply correlated
with each other by the net-quark fraction $z$. 

\begin{figure}
\includegraphics[scale=0.42]{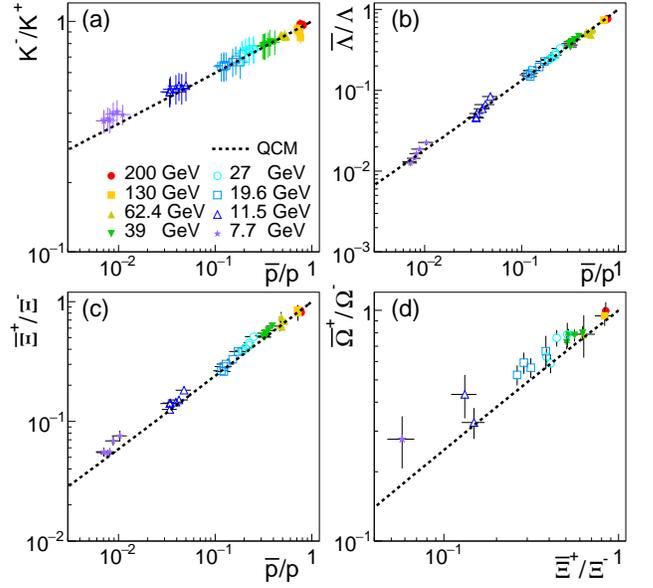}\caption{Anti-hadron to hadron yield ratios in Au+Au collisions at different
collision energies. Symbols are experimental data in central and semi-central
(centrality <60\%) collisions \cite{Adcox:2002au,Adams:2003fy,Adams:2006ke,Abelev:2008ab,Arsene:2009jg,Aggarwal:2010ig,Adamczyk:2017iwn,Adam:2019koz}.
Dashed lines are model results. }
\label{fig:dndyCorr}
\end{figure}

Based on Eqs. (\ref{eq:rppbar}-\ref{eq:rOmg}), we can build several
multi-hadron correlations as more sensitive tests of quark combination
mechanism. In Fig. \ref{fig:dndyCorr} (a), we firstly show the correlation
between $K^{-}/K^{+}$ and $\bar{p}/p$. This correlation shows how
the production of baryon and that of meson in heavy-ion collisions
are simultaneously influenced by the baryon quantum number density
characterized by net-quark fraction $z$ in our model. Symbols are
experimental data at mid-rapidity in central and semi-central collisions
\cite{Adcox:2002au,Adams:2003fy,Adams:2006ke,Abelev:2008ab,Arsene:2009jg,Aggarwal:2010ig,Adamczyk:2017iwn,Adam:2019koz}.
Error bars are the quadratic sum of statistical and systematic uncertainties.
The dashed line is the result of QCM by Eqs. (\ref{eq:rppbar}) and
(\ref{eq:rKpm}). Different from our previous work \cite{Song:2013isa}
and previous experimental measurements \cite{Arsene:2009jg}, here
we show the correlation in double-logarithmic coordinates in order
to provide a full and clear presentation because ratio $\bar{p}/p$
changes much faster than $K^{-}/K^{+}$. We see that experimental
data in double-logarithmic coordinates behave as almost a straight
line and our model can well reproduce this correlation. 

In Fig. \ref{fig:dndyCorr} (b) we show the correlation between $\bar{\Lambda}/\Lambda$
and $\bar{p}/p$. We see that experimental data, symbols in the figure,
exhibit a linear behavior in double-logarithmic coordinates. This
behavior can be perfectly reproduced in our model by Eqs. (\ref{eq:rppbar})
and (\ref{eq:rLam}), see the dashed line. In addition, we note that
ratio $\bar{\Lambda}/\Lambda$ decreases slower than $\bar{p}/p$
by a factor $(1-z)/(1+z)=N_{\bar{q}}/N_{q}$. This is because that,
compared with $\bar{p}/p$, ratio $\bar{\Lambda}/\Lambda$ involves
a strangeness neutralization $N_{\bar{s}}/N_{s}=1$

In Fig. \ref{fig:dndyCorr} (c) we show the correlation between $\bar{\Xi}^{+}/\Xi^{-}$
and $\bar{p}/p$. Experimental data also exhibit a linear behavior
in double-logarithmic coordinates. Since $\bar{\Xi}^{+}/\Xi^{-}$
involves double effect of strangeness neutralization $(N_{\bar{s}}/N_{s})^{2}=1$,
$\bar{\Xi}^{+}/\Xi^{-}$ decreases slower than $\bar{\Lambda}/\Lambda$
as the function of $\bar{p}/p$. Our model result Eqs. (\ref{eq:rppbar})
and (\ref{eq:rXi}), the dashed line in the figure, can well describe
data at $\sqrt{s_{NN}}\geq$ 11.5 GeV and slightly under-estimates
$\bar{\Xi}^{+}/\Xi^{-}$ at $\sqrt{s_{NN}}=$ 7.7 GeV. 

In Fig. \ref{fig:dndyCorr} (d), we further show the correlation among mutl-strangeness hadrons 
$\bar{\Omega}^{+}/\Omega^{-}$ and $\bar{\Xi}^{+}/\Xi^{-}$. Experimental
data in double-logarithmic coordinates also exhibit a linear behavior.
Because $\Omega^{-}$( $\bar{\Omega}^{+}$) completely consists of
strange (anti)quarks, ratio $\bar{\Omega}^{+}/\Omega^{-}$ involves
triple effect of strangeness neutralization and therefore it decreases
slower than $\bar{\Xi}^{+}/\Xi^{-}$. The model result by Eqs. (\ref{eq:rXi})
and (\ref{eq:rOmg}), the dashed line in the figure, can roughly describe
data at $\sqrt{s_{NN}}\geq$ 11.5 GeV and under-estimates $\bar{\Omega}^{+}/\Omega^{-}$
at $\sqrt{s_{NN}}=$ 7.7 GeV to a certain extent. 

Some discussions on above results are necessary. 
First, we emphasize the key physics in our model relating to multi-hadron
correlations shown in Fig. \ref{fig:dndyCorr}. As shown by Eqs. (\ref{eq:rppbar})-(\ref{eq:rOmg}),
correlations among yield ratios of anti-hadron to hadron are not sensitive
to absolute numbers of quarks and antiquarks at hadronization and
non-perturbative parameters $R_{V/P}$ and $R_{D/O}$ introduced in
the model. Therefore, these yield correlations are only related to
two basic features of quark combination in our model. (1) free combination.
Newborn quarks and antiquarks, net-quarks are all treated as individual
(anti-)quarks and freely take part in combination. (2) flavor independent
combination probability. We take $\overline{N}_{B}/N_{q}^{3}$ as
the averaged probability of $q_{1}q_{2}q_{3}$ forming a baryon and
$\overline{N}_{M}/(N_{q}N_{\bar{q}})$ as the averaged probability
of $q_{1}\bar{q}_{2}$ forming a meson. No sophisticated flavor correction
is made at the moment. From Fig. \ref{fig:dndyCorr}, we see that
such a global quark combination model provides a systematic description
on multi-hadron yield correlations in Au+Au collisions, at least at
$\sqrt{s_{NN}}\geq$ 11.5 GeV. Therefore, this is a clear signal of
quark combination mechanism at hadronization in these collisions. 

Second, the comparison between our model calculation and experimental
data in Au+Au collisions at $\sqrt{s_{NN}}=$ 7.7 GeV may indicate
some physics beyond the key features of quark combination discussed
above. As $\bar{p}/p$ and $\bar{\Lambda}/\Lambda$ ratios are reproduced,
we see that model results for $K^{-}/K^{+}$,$\bar{\Xi}^{+}/\Xi^{-}$and
$\bar{\Omega}^{+}/\Omega^{-}$ are smaller than experimental data
to a certain extent. This may be related to the point (1) discussed
above. In Au+Au collisions at low energy, the colliding nucleons may
not break completely. A part of nucleon fragments may do not behave
as the individual up/down quarks and freely take part in the combination
with newborn quarks and antiquarks; Instead, they behave as diquarks
and can form proton by combining with a up/down quark or form $\Lambda$
by combining with a strange quark. Because these net-quarks only contribute
to proton and $\Lambda$ production, net-quark fraction $z$ used
in Eqs. (\ref{eq:rKpm}), (\ref{eq:rXi}) and (\ref{eq:rOmg}) for
$K^{-}/K^{+}$, $\bar{\Xi}^{+}/\Xi^{-}$and $\bar{\Omega}^{+}/\Omega^{-}$
should be smaller than that in $\bar{p}/p$ and $\bar{\Lambda}/\Lambda$.
This consideration can increase ratios $K^{-}/K^{+}$, $\bar{\Xi}^{+}/\Xi^{-}$and
$\bar{\Omega}^{+}/\Omega^{-}$ at the given $\bar{p}/p$ and $\bar{\Lambda}/\Lambda$
ratios and therefore can qualitatively improve the description at
$\sqrt{s_{NN}}=$ 7.7 GeV in the current model. Such a sophisticated
effect of net-quarks is worthwhile to be studied in detail in the
future works.

\section{properties of quark distributions \label{sec:quark_prop}}

By studying experimental data of hadronic $p_{T}$ spectra, we have obtained $p_{T}$ spectra of quarks at hadronization, which are shown in Fig. \ref{fig:fuds}. In this section, we study properties of these obtained quark distributions at hadronization. 

\begin{figure}[h]
\includegraphics[width=0.85\linewidth]{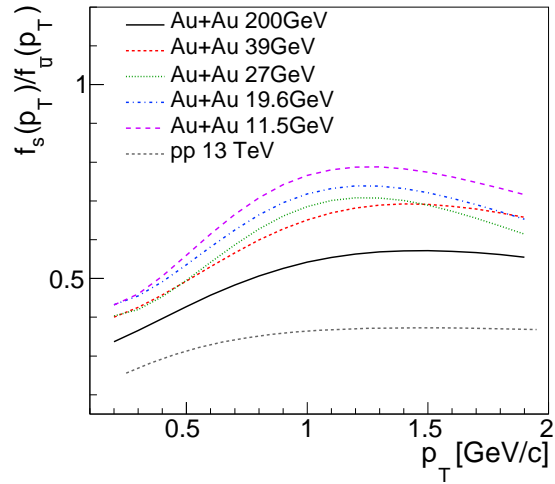}\caption{Spectrum ratio $f_{s}\left(p_{T}\right)/f_{\bar{u}}\left(p_{T}\right)$ in central Au+Au collisions at $\sqrt{s_{NN}}=$ 200, 39, 27, 19.6 and 11.5 GeV, and that in $pp$ collisions at $\sqrt{s}=13$ TeV.}
\label{fig:rsu_pt}
\end{figure}

We firstly calculate the ratio $f_{s}\left(p_{T}\right)/f_{\bar{u}}\left(p_{T}\right)$
and study its $p_{T}$ dependence. Fig. \ref{fig:rsu_pt} shows results
in central Au+Au collisions at $\sqrt{s_{NN}}=$ 200, 39, 27, 19.6 and 11.5 GeV. 
Result of $f_{s}\left(p_{T}\right)/f_{\bar{u}}\left(p_{T}\right)$ in $pp$ collisions at $\sqrt{s}=13$ TeV \cite{Zhang:2018vyr} is also presented. We see that ratios in Au+Au collisions are obviously higher than that in $pp$ collisions. This means that the production of strange quarks in the studied $p_T$ range in Au+Au collisions is significantly enhanced. In addition, we see that ratios in Au+Au collisions at these collision energies in the low $p_{T}$ range ($p_{T}\lesssim1.3$ GeV/c) all increase with $p_{T}$. It is similar to $pp$ results. This property is related to the complex (non-)perturbative QCD evolution in connection with quark masses in partonic phase.

We note that the increase of the ratio at low $p_{T}$ in heavy-ion collisions can be qualitatively understood by thermal statistics. In the case of thermal equilibrium such as the Boltzmann distribution $dN/(p_{T}dp_{T})\propto\exp\left(-\sqrt{p_{T}^{2}+m^{2}}/T\right)$ in two-dimensional transverse momentum space, heavier mass will lead to flatter shape of the $p_{T}$ distribution and thus lead to the increase of the ratio. 
However, Boltzmann distribution at hadronization temperature in the rest
frame can not directly describe the obtained quark distributions in
Fig. \ref{fig:fuds}. We should further take into account the contribution
of the collective radial flow in the prior partonic phase evolution
in heavy-ion collisions. As an illustration, we consider a simple
situation, that is, Boltzmann distribution in the two-dimensional
transverse momentum space boosted with a radial flow velocity $v_{\perp}$.
In this case the distribution is
\begin{equation}
\frac{dN}{p_{T}dp_{T}}\propto\frac{1}{E}E^{*}\left(v_{\perp}\right)\exp[-E^{*}\left(v_{\perp}\right)/T]\label{eq:BoltzFit}
\end{equation}
with $E^{*}\left(v_{\perp}\right)=\gamma_{\perp}\left(E-v_{\perp}p_{T}\right)$
and $E=\sqrt{p_{T}^{2}+m^{2}}$. If we assume that quarks of different
flavors at hadronization are thermalized and boosted with the same
radial velocity, we can use the above formula to simultaneously describe
the extracted $f_{\bar{u}}\left(p_{T}\right)$ and $f_{s}\left(p_{T}\right)$
in the low $p_{T}$ range ($p_{T}\leq1.3$ GeV/c) in Fig. \ref{fig:fuds}.
According to our previous work \cite{Song:2019sez}, the hadronization
temperature is taken as $T=$ (0.164, 0.163, 0.162, 0.161, 0.156)
GeV in central Au+Au collisions at $\sqrt{s_{NN}}=$ (200, 39, 27,
19.6, 11.5) GeV, respectively. We obtain radial flow velocity $v_{\perp}/c\approx$
(0.39, 0.28, 0.27, 0.25, 0.23) at these collision energies. 

\begin{figure}[h]
\includegraphics[width=\linewidth]{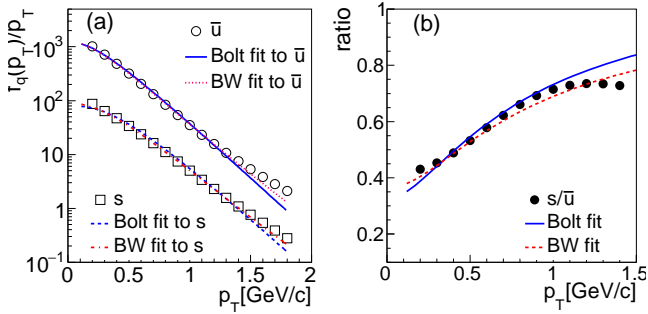}
    \caption{(a) Fit results for quark $p_{T}$ spectra at hadronization in central
Au+Au collisions at $\sqrt{s_{NN}}=$ 19.6 GeV by Boltzmann formula
Eq. (\ref{eq:BoltzFit}) and by blast-wave model at the same hadronization
temperature 0.161 GeV and radial flow velocity 0.25 $c$. (b) ratio
$f_{s}\left(p_{T}\right)/f_{\bar{u}}\left(p_{T}\right)$ by two fit
methods. Symbols are quark $p_{T}$ spectra and lines are fit results.}

\label{fig:fqpt_fits}
\end{figure}

We find that these results for radial flow velocity are consistent
with our previous extraction by a hydrodynamics-motivated blast-wave
model \cite{Schnedermann:1993ws} fit of $f_{s}\left(p_{T}\right)$
with the same hadronization temperature \cite{Song:2019sez}. We
also find that $f_{\bar{u}}\left(p_{T}\right)$ and $f_{s}\left(p_{T}\right)$
extracted in this work can also be consistently described in blast-wave
mode. Here, taking central Au+Au collisions at $\sqrt{s_{NN}}=$ 19.6
GeV as an example, we show in Fig. \ref{fig:fqpt_fits} (a) the fit
results of quark $p_{T}$ spectra by Eq. (\ref{eq:BoltzFit}) and
those by blast-wave model at the same hadronization temperature 0.161
GeV and radial flow velocity 0.25 $c$. We see that two fit methods
give the consistent description on quark $p_{T}$ spectra. In addition,
we see from Fig. \ref{fig:fqpt_fits} (b) that the increase part of
the ratio $f_{s}\left(p_{T}\right)/f_{\bar{u}}\left(p_{T}\right)$
in the low $p_{T}$ range can be reasonably described. 

\begin{figure}
\includegraphics[scale=0.35]{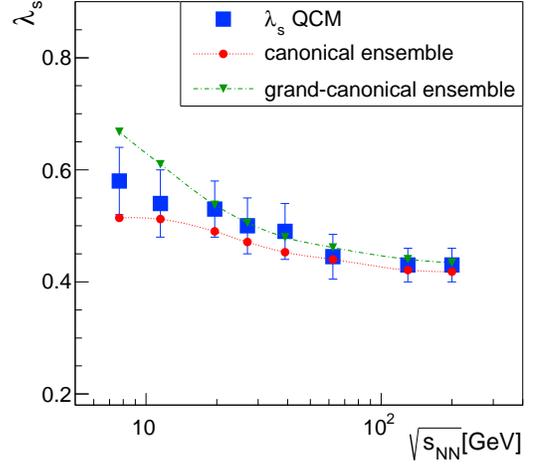}\caption{Strangeness factor $\lambda_{s}=N_{s}/N_{\bar{u}}$ in central Au+Au
collisions at different collision energies.}

\label{fig:ls_snn}
\end{figure}

Fig. \ref{fig:rsu_pt} also shows the ratio $f_{s}\left(p_{T}\right)/f_{\bar{u}}\left(p_{T}\right)$
globally increases with the decrease of collision energies. To study
this energy dependence of strangeness, we calculate the strangeness
factor 
\begin{equation}
\lambda_{s}=\frac{N_{s}}{N_{\bar{u}}}
\end{equation}
and present results in Fig. \ref{fig:ls_snn}. Here, results of $\lambda_{s}$
in central Au+Au collisions at $\sqrt{s_{NN}}=$ 62.4 and 130 GeV
are also shown. The uncertainty of $\lambda_{s}$ is caused by that
of experimental data for yield ratios such as $K/\pi$, $\Lambda/p$,
$\bar{\Lambda}/\bar{p}$. We note that these new results of $\lambda_{s}$
are consistent with our previous works \cite{Shao:2009uk,Sun:2011kj,Shao:2015rra}.
Compared with $\lambda_{s}\approx0.3$ in elementary collisions such
as $e^{+}e^{-}$ and $pp$ reactions, we see that $\lambda_{s}\gtrsim0.42$
in heavy-ion collisions is obviously enhanced. We also see that $\lambda_{s}$
increases as the decrease of collision energy.

The dependence of $\lambda_{s}$ on collision energy is related to
the varied baryon quantum number density. In this paper, we study
this energy dependence in the framework of thermodynamics for  quark
system. We consider a thermal system consisting of constituent
quarks and antiquarks. In our quark combination model, constituent
quarks and antiquarks are regarded as the effective degrees of freedom
at hadronization and they freely combine into baryons and/or mesons
at hadronization. Therefore, we can treat such a constituent quark
system as the classical gas.

Because we study the production of hadrons in mid-rapidity range,
we firstly discuss the case of grand-canonical ensemble. The number
of quark under Fermi-Dirac statistics in the grand-canonical ensemble
is 
\begin{align}
N_{f} & =g\int\frac{d^{3}xd^{3}p}{\left(2\pi\right)^{3}}\frac{1}{\exp\left[\left(E-\mu_{f}\right)/T\right]+1}\nonumber \\
 & =g\frac{Vm^{2}T}{2\pi^{2}}\sum_{n=1}\frac{\left(-1\right)^{n+1}}{n}K_{2}\left(n\frac{m}{T}\right)\exp\left(n\frac{\mu_{f}}{T}\right),\label{eq:Nqi_gce}
\end{align}
where $V$ is system volume and $T$ is temperature. $g=6$ is degeneracy
factor of quark, $m$ is quark mass and $\mu_{f}$ is chemical potential
of quark $f$. $K_{2}$ is the modified Bessel function of the second
kind. As discussions of local strangeness conservation in Sec. \ref{sec:Sconservation},
we set $\mu_{s}=\mu_{\bar{s}}=0$ to get $N_{s}=N_{\bar{s}}$. We
assume the iso-spin symmetry between up and down quarks, then we have
$\mu_{u}=\mu_{d}=-\mu_{\bar{u}}=-\mu_{\bar{d}}=\mu_{B}/3$. Then,
the strangeness factor in grand-canonical ensemble (GCE) of free quark
gas is 
\begin{align}
\lambda_{s}^{\left(GCE\right)} & =\frac{m_{s}^{2}\sum_{n=1}\frac{\left(-1\right)^{n+1}}{n}K_{2}\left(n\frac{m_{s}}{T}\right)}{m_{u}^{2}\sum_{n=1}\frac{\left(-1\right)^{n+1}}{n}K_{2}\left(n\frac{m_{u}}{T}\right)\exp[-n\frac{\mu_{B}}{3T}]}\nonumber \\
 & \approx\lambda_{s,\mu_{B}=0}^{\left(GCE\right)}\exp[\frac{\mu_{B}}{3T}].\label{eq:ls_qce}
\end{align}
In the second line, we consider only the leading term $n=1$, which
is corresponding to the Boltzmann statistics. As we know, baryon chemical
potential is increased as the decrease of the collision energy. Therefore,
the collision energy dependence of $\lambda_{s}$ is qualitatively
understood. 

For a demonstrative calculation for the collision energy dependence
of $\lambda_{s}^{\left(GCE\right)}$, we first apply Eq. (\ref{eq:ls_qce})
with the re-tuned $m_{u}=0.3$ GeV and $m_{s}=0.54$ GeV to give a
proper strangeness at vanishing baryon chemical potential $\lambda_{s,\mu_{B}=0}^{\left(GCE\right)}\approx0.42$.
Then, we apply Eq. (\ref{eq:Nqi_gce}) to fit the total quark number
$x=N_{q}+N_{\bar{q}}$ and net-quark number $xz=N_{q}-N_{\bar{q}}$
integrated from Fig. \ref{fig:fuds} and obtain $V$ and $\mu_{B}$
of quark system at hadronization. Here, the hadronization temperature
is taken as before. Then we substitute $\mu_{B}$ and $T$ into Eq.
(\ref{eq:ls_qce}). The calculated results for $\lambda_{s}^{\left(GCE\right)}$
at RHIC energies are shown as triangles with the dashed line in Fig.
\ref{fig:ls_snn}. We see that the extracted $\lambda_{s}$ in quark
combination model can be explained by grand-canonical ensemble of
quark gas system at $\sqrt{s_{NN}}\gtrsim20$ GeV. At lower two collision
energies, GCE results are higher than our extraction. 

Considering the longitudinal rapidity space of heavy-ion collisions
is finite, in particular, $y_{beam}<2.5$ at $\sqrt{s_{NN}}\leq11.5$
GeV, the studied mid-rapidity range $|y|<0.5$ is not a tiny fraction
of the whole system, the hadron production in the midrapidity range
may be influenced by effects of global charge conservation not only
strangeness conservation but also baryon quantum number conservation.
Therefore, grand-canonical treatment may be not perfectly suitable.
We now consider the result of canonical ensemble for free quark system.
We apply the method of canonical statistics \cite{Becattini:2004rq}
to obtain the property of quark number distribution under the constraint
of finite baryon quantum number and strangeness. We put the detailed
derivation into Appendix \ref{sec:CE_Q}. The inputs of canonical
ensemble are volume $V$, temperature $T$, and charges ($B$,$Q$,$S$).
The temperature is set to the hadronization temperature whose values
at different collision energies are taken as before. As discussions
in Sec. \ref{sec:Sconservation}, we take $S=0$. $V$, $B$ and $Q$
can be determined by fitting the quark and antiquark numbers integrated
from Fig. \ref{fig:fuds}. Results of $\lambda_{s}^{\left(CE\right)}$
for canonical ensemble of quark system are shown as solid circles
with the dotted line in Fig. \ref{fig:ls_snn}. We see that $\lambda_{s}^{\left(CE\right)}$
is also increased with the decrease of collision energy and is smaller
than $\lambda_{s}^{\left(GCE\right)}$, in particular, at low collision
energies. Our extracted $\lambda_{s}$ is roughly located in the middle
of two ensembles. 

\section{Summary and discussions \label{sec:Summary}}

In this paper, we have applied a quark combination model with equal-velocity
combination approximation to systematically study the production of
hadrons in Au+Au collisions at RHIC energies. The model applied in
this work is motivated by our recent findings for the constituent
quark number scaling property of hadronic $p_{T}$ spectra in high
energy $pp$, $p$A and AA collisions \cite{Song:2017gcz,Zhang:2018vyr,Song:2019sez}. After systematic study of $p_{T}$
spectra and yields for hadrons, we found that our quark combination
model provides a good explanation on the experimental data in Au+Au
collisions at $\sqrt{s_{NN}}\geq11.5$ GeV. This suggests that the
constituent quark degrees of freedom still play important role even
at low RHIC energy, which is closely related to the deconfinement
at these collision energies. 

By study of hadronic $p_{T}$ spectra and yields in these collisions,
we obtained $p_{T}$ spectra and numbers of constituent quarks and
antiquarks at hadronization. We calculated the net strangeness $N_{\bar{s}}-N_{s}$
at mid-rapidity and showed the strangeness neutralization is well
satisfied at mid-rapidity in heavy-ion collisions at RHIC energies.
We studied the spectrum ratio of strange quarks to newborn up/down
quarks $f_{s}\left(p_{T}\right)/f_{\bar{u}}\left(p_{T}\right)$ and
the $p_{T}$ integrated number ratio $\lambda_{s}=N_{s}/N_{\bar{u}}$
at different collision energies. We applied the basic thermal statistics
for free constituent quark system to understand these properties of
strange quarks relative to up/down quarks. 

We emphasize the key physics in our model which are responsible for
successfully explaining experimental data of hadronic $p_{T}$ spectra
and yields. First, the model takes the constituent quarks and antiquarks
as the effective interface connecting the strongly-interacting system
before hadronization and that after hadronization. We assume the constituent
quarks and antiquarks as effective degrees of freedom for the strongly-interacting
quark-gluon system just before hadronization. On the other hand, we
take the constituent quark model to describe the static structure
of hadrons in the ground state. In this scenario, the equal-velocity
combination of these constituent quarks and antiquarks is a reasonable
approximation and is indeed supported by the quark number scaling
property for $p_{T}$ spectra of hadrons observed in experiments \cite{Song:2017gcz,Zhang:2018vyr,Song:2019sez}.
The study in this paper further showed that the equal-velocity combination
can systematically describe the production of different kinds of hadrons
in heavy-ion collisions at RHIC energies. Second, the model includes
reasonable considerations for the unity of hadronization and the linear
response property, see detailed discussions in \cite{Song:2013isa}.
This is very important to reproduce the multi-hadron yield correlations
shown in Fig. \ref{fig:dndyCorr}. For example, in a naive inclusive
view of combination, we have $N_{\Omega}\propto N_{s}^{3}$ and $N_{\bar{\Omega}}\propto N_{\bar{s}}^{3}$
and therefore $\bar{\Omega}^{+}/\Omega^{-}\propto\left(N_{\bar{s}}/N_{s}\right)^{3}\approx1$
which is independent of collision energy. However, yield of $\Omega^{-}$
in our model is not only dependent on $N_{s}^{3}$ but also dependent
on the global system information shown as in Eq. (\ref{eq:PBj});
thus we have $\bar{\Omega}^{+}/\Omega^{-}=\left[(1-z)/(1+z)\right]^{a-3}$
in Eq. (\ref{eq:rOmg}) which decreases with the decrease of collision
energy. 

In order to further test our EVC model, the following two aspects are deserved to study in depth in future works. The first is the effect of momentum correlations among quarks and antiquarks at hadronization.  It is neglected in this paper, see Eqs.~(\ref{eq:fqqq}) and (\ref{eq:fqqbar}). In general, quarks and antiquarks at hadronization always have some momentum correlations. In heavy-ion collisions, the collective flow formed in parton phase will cause a certain correlation among momenta of quarks and antiquarks. This correlation not only influences the inclusive momentum spectra of hadrons to a certain extent but also influences multi-particle momentum correlations more directly. In the future work, we will study this effect by building sensitive physical observables in EVC model. 
The second is the production of short-lived resonances. In heavy-ion collisions, yield and momentum spectra of finally-observed resonances such as $K^{*}(892)$ are strongly influenced by re-scatterings among hadrons. In the future work, we will systematically consider this hadronic rescattering effect and study the production mechanism of the short-lived resonances at hadronization.   

\section{Acknowledgments}

This work is supported in part by the National Natural Science Foundation
of China under Grant No. 11975011, Shandong Province Natural Science
Foundation under Grant Nos. ZR2019YQ06 and ZR2019MA053, and Higher
Educational Youth Innovation Science and Technology Program of Shandong
Province (2019KJJ010, 2020KJJ004).

\appendix

\section{Quark number distribution in canonical ensemble \label{sec:CE_Q}}

The probability of a single state in the canonical ensemble is

\begin{equation}
P_{state}=\frac{1}{Z\left(\bm{Q}\right)}e^{-E_{state}\beta}\delta_{\bm{Q},\bm{Q}_{state}},\label{eq:Pstate}
\end{equation}
where $\beta=1/T$ is the inverse temperature, $E_{state}$ is the
energy of the state and $\bm{Q}_{state}$ are abelian charges of the
state

\begin{eqnarray}
\mathbf{Q}_{state} & = & \sum_{j=1}^{K}\mathbf{q}_{j}N_{j}.
\end{eqnarray}
Here $N_{j}$ is the number of particle $j$ in the current state
and $\mathbf{q}_{j}=\left(B_{j},Q_{j},S_{j}\right)$ is the quantum
number vector for the $j$-th particle. $K$ is the number of particle
species.

The multiplicity distribution of particles can be obtained from the
generating function associated to the canonical partition function
$Z(\bm{Q})$ \cite{Becattini:2004rq} and is expressed as

\begin{align}
 & P\left(\left\{ N_{j}\right\} \right)\label{eq:PNj_becattini}\\
 & =\frac{1}{Z\left(\bm{Q}\right)}\,\prod_{j=1}^{K}\left\{ \sum_{\left\{ h_{n_{j}}\right\} }\prod_{n_{j}=1}^{N_{j}}\left[\frac{\left[z_{j}\left(n_{j}\right)\right]^{h_{n_{j}}}}{n_{j}^{h_{n_{j}}}h_{n_{j}}!}\right]\right\} \delta_{\bm{Q},\sum_{j}\bm{q_{j}}N_{j}},\nonumber 
\end{align}
where the summation takes different configurations for $\left\{ h_{n_{j}}\right\} $
into account under the condition $\sum_{n_{j}=1}^{\infty}n_{j}h_{n_{j}}=N_{j}$.
\begin{align}
z_{j}\left(n_{j}\right) & =\left(\mp\right)^{n_{j}+1}\frac{gV}{\left(2\pi\right)^{3}}\int d^{3}p\,e^{-n\beta E}\nonumber \\
 & =\left(\mp\right)^{n_{j}+1}gV\frac{m^{2}}{2\pi^{2}n_{j}\beta}K_{2}(n_{j}\beta m).
\end{align}

Now, we consider the thermal system consisting of constituent quarks
and antiquarks. In our quark combination model, constituent quarks
and antiquarks are regarded as the effective degrees of freedom at
hadronization and they freely combine to form baryons and/or mesons
at hadronization. Therefore, we can simply apply above formula to
obtain the number distribution of these ``free'' constituent quarks
and antiquarks under canonical statistics. Here, we consider up, down,
strange quarks and their antiparticles. Index $j$ denotes $u,d,s$,
$\bar{u},\bar{d},\bar{s}$ and $K=6$. 

Eq. (\ref{eq:PNj_becattini}) can be further denoted as 

\begin{align}
 & P\left(\left\{ N_{j}\right\} \right)=\frac{1}{Z\left(\bm{Q}\right)}\left[\prod_{j=1}^{K}\mathcal{Z}\left(S_{N_{j}}\right)\right]\delta_{\bm{Q},\sum_{j}N_{j}\bm{q_{j}}},
\end{align}
where $\mathcal{Z}\left(S_{N_{j}}\right)$ is cycle-index polynomial
of symmetric group. $\mathcal{Z}\left(S_{N_{j}}\right)$ can be numerically
evaluated using the recurrence relation 
\begin{equation}
\mathcal{Z}\left(S_{N}\right)=\frac{1}{N}\sum_{l=1}^{N}z\left(l\right)\mathcal{Z}\left(S_{N-l}\right)
\end{equation}
with $\mathcal{Z}\left(S_{0}\right)=1$ and $\mathcal{Z}\left(S_{1}\right)=z\left(1\right)$. 

As discussed in Sec. \ref{sec:Sconservation}, we take strangeness
neutralization $S=0$ and therefore $N_{s}=N_{\bar{s}}$. For constraints
of baryon charge $B$ and electric charge $Q$, we denote them as
$N_{u}-N_{\bar{u}}=B+Q$ and $N_{d}-N_{\bar{d}}=2B-Q$. Finally the
joint distribution function of quark numbers and antiquark numbers
is 
\begin{align}
 & P\left(N_{d},N_{u},N_{s},N_{\bar{d}},N_{\bar{u}},N_{\bar{s}}\right)\nonumber \\
 & =\frac{1}{Z\left(\bm{Q}\right)}\mathcal{Z}\left(S_{N_{d}}\right)\mathcal{Z}\left(S_{N_{\bar{d}}}\right)\,\delta_{N_{d},N_{\bar{d}}+2B-Q}\nonumber \\
 & \,\,\,\,\,\times\mathcal{Z}\left(S_{N_{u}}\right)\mathcal{Z}\left(S_{N_{\bar{u}}}\right)\delta_{N_{u},N_{\bar{u}}+B+Q}\nonumber \\
 & \,\,\,\,\,\times\mathcal{Z}\left(S_{N_{s}}\right)\mathcal{Z}\left(S_{N_{\bar{s}}}\right)\delta_{N_{s},N_{\bar{s}}}.
\end{align}
We obtain the averaged number of quarks by 
\begin{align}
\overline{N}_{s} & =\sum_{\left\{ N_{q_{i}}\right\} }N_{s}P\left(N_{d},N_{u},N_{s},N_{\bar{d}},N_{\bar{u}},N_{\bar{s}}\right),\\
\overline{N}_{\bar{u}} & =\sum_{\left\{ N_{q_{i}}\right\} }N_{\bar{u}}P\left(N_{d},N_{u},N_{s},N_{\bar{d}},N_{\bar{u}},N_{\bar{s}}\right),
\end{align}
and calculate strangeness factor $\lambda_{s}^{\left(CE\right)}=\overline{N}_{s}/\overline{N}_{\bar{u}}$
in canonical ensemble of free quark system. 

\bibliographystyle{apsrev4-1}
\bibliography{ref}

\end{document}